\documentclass[aps,prd,onecolumn,groupedaddress]{revtex4}
\usepackage{graphicx}
\usepackage{subfigure}
\usepackage{dcolumn}
\usepackage{bm}
\usepackage{color}
\usepackage{longtable}
\usepackage{supertabular}
\usepackage[T1]{fontenc}
\usepackage{epstopdf}
\usepackage{amsmath}
\usepackage{float}
\usepackage{hyperref}

\makeatletter

\newcommand{\Rmnum}[1]{\expandafter\@slowromancap\romannumeral #1@}
\makeatother

\begin{document}
\title{Electrostatic effect due to patch potentials between closely spaced surfaces}
\author{Jun Ke$^{1}$}
\author{Wen-Can Dong$^{2}$}
\author{Sheng-Hua Huang$^{1}$}
\author{Yu-Jie Tan$^{2}$}
\author{Wen-Hai Tan$^{3}$}
\author{Shan-Qing Yang$^{3}$}
\author{Cheng-Gang Shao$^{2}$}\email[E-mail: ]{cgshao@hust.edu.cn}
\author{Jie Luo$^{1}$}\email[E-mail: ]{luojiethanks@126.com}

\affiliation{
$^{1}$ School of Mechanical Engineering and Electronic Information, China University of Geosciences, Wuhan 430074, People's Republic of China\\
$^{2}$ MOE Key Laboratory of Fundamental Physical Quantities Measurement, Hubei Key Laboratory of Gravitation and Quantum Physics, PGMF and School of Physics, Huazhong University of Science and Technology, Wuhan 430074,  People's Republic of China\\
$^{3}$TianQin Research Center for Gravitational Physics and School of Physics and Astronomy, Sun Yat-sen University (Zhuhai Campus), Zhuhai 519082, People's Republic of China
}

\date{\today}
\begin{abstract}
The spatial variation and temporal variation in surface potential are important error sources in various precision experiments and deserved to be considered carefully. In the former case, the theoretical analysis shows that this effect depends on the surface potentials through their spatial autocorrelation functions. By making some modification to the quasi-local correlation model, we obtain a rigorous formula for the patch force, where the magnitude is proportional to ${\frac{1}{{{a}^{2}}}{{(\frac{a}{w})}^{\beta (a/w)+2}}}$ with ${a}$ the distance between two parallel plates, ${w}$ the mean patch size, and ${\beta}$ the scaling coefficient from ${-2}$ to ${-4}$. A torsion balance experiment is then conducted, and obtain a 0.4 mm effective patch size and 20 mV potential variance. In the latter case, we apply an adatom diffusion model to describe this mechanism and predicts a ${f^{-3/4}}$ frequency dependence above 0.01 ${\rm mHz}$. This prediction meets well with a typical experimental results. Finally, we apply these models to analyze the patch effect for two typical experiments. Our analysis will help to investigate the properties of surface potentials.
\end{abstract}

\maketitle


\section{INTRODUCTION}\label{section1}
The surface of an ideal metallic conductor is often assumed to be an equipotential. However, it would not be true for real metallic surface according to the measurements of surface potential \cite{1,2,3,4,5,6,7,8,9,10}, for example, Camp ${et \, al}$ measured surface potential variations of 70 mV over scales of 10 mm in a mental sample \cite{2}. Furthermore, fluctuations in the electric potential are also be observed by different experiments \cite{4,9,10}. This phenomenon of the spatial variations and temporal fluctuations in surface potential is usually referred to as the ``patch effect'' \cite{11}. Patch potentials can be generated by many reasons, such as the regions of different crystal orientation, the nonuniform segregation of the elements and the presence of contaminants adsorbed on the surface \cite{1,3,4,5,7}. These dynamic patches produce an electrostatic force or electric filed noise which is different from the equipotential situation \cite{11,12}. Since the magnitude of this noise is depended on the specific distribution and temporal fluctuations of surface potentials, its influence is difficult to evaluate with high precision. This electrostatic noise has been recognized as an important error source in various precision experiments, including tests of the gravitational inverse-square-law (ISL) \cite{13,14,15,16,17}, measurement of the Casimir force \cite{18,19,20}, heating in ion traps \cite{21,22,23,24}, spaceborne gravitational wave detection \cite{25,26,27,28,29}, and so on \cite{30,31}. For example, the measurement precision of the GP-B mission and ISL experiments at short range were mainly limited by the patch effect \cite{13,14,15,16,31,32}. In addition, the coupling between the potential fluctuation with net free charge on the test mass is also a major acceleration noise term for the Laser Interferometer Space Antenna (LISA) \cite{29,34,35,36,37}. Therefore, it is significant to investigate the properties of surface potential in pursuit of higher experimental sensitivity.

Generally, the methods to study the origin and influence of patch potential includes the experimental measurement and the theoretical modeling. For the experimental measurement of the patch effect, the Kelvin probe and the torsion pendulum methods are often used \cite{4,5,6,7,8,9,32}. The Kelvin probe force microscopy (KPFM) can measure the potentials over the testing region with an extremely high spatial resolution of about several nanometers. However, the potential resolution of KPFM is limited to about 1 ${\rm mV/Hz^{1/2}}$ when the probe tip remained at the same place and measurements \cite{4}. The torsion pendulum can measure the temporal fluctuations of surface potentials with a resolution of 30 ${\rm \mu V/Hz^{1/2}}$, but it can not obtain the information about the spatial potential distribution \cite{9}. Therefore, Huazhong University of Science and Technology (HUST) research group developed a torsion pendulum with a scanning probe to measure the surface potential \cite{6,7}. Their results show that the voltage resolution can reach a level of 4 ${\rm \mu V/Hz^{1/2}}$ for a millimetre area and the spatial distribution is about 330 ${\rm \mu V}$ at 0.125 mm spatial resolution. However, the above measurements did not show a consistent pattern, which leads to the physical origin of the patch potentials still remains mysterious so far.

The theoretical modeling itself can provide valuable predictions for the influences of patch potentials \cite{24}. Currently, the theoretical analysis of the patch effect in precision experiments focused on two main problems, namely, the electrostatic force between closed spaced metallic surfaces \cite{11,12,13,14,15,16,17,18,19,20,38} and the electric filed noise above a conducting surface \cite{21,22,23,24}. In the former case, Speake ${et \, al}$ obtained a rigorous formula for the patch force between two infinite parallel plates by solving the Laplace's equation, and also can be extended to the sphere-plane geometry\cite{11,12,13}. Their results show that this kind of force or force power spectrum depends on the potentials only through their spatial autocorrelation functions (SAFs). Subsequently, a number of correlation models were proposed to describe the spatial distribution of patch potentials, such as the sharp-cutoff model \cite{12}, the quasi-local correlation model \cite{19} and the exponential model \cite{21}. Based on these models, the magnitude of spatial patch noise can be estimated to some extent. In the latter case, Dubessy ${et \, al}$ analyzed the time-dependent electric noise above a conducting surface \cite{21}. Under the assumption of the temporal and spatial variations of the patches decouple, this noise is also depends on the potentials only through their SAFs. Therefore, similar analyses are conducted for estimating this noise \cite{22,23,24}. In those cases, the SAFs are all the information we have about the spatial patch potentials. While the spatial patch potentials on metal surfaces are relatively well understood, little is known about their fluctuations. Although the fluctuating adatomic dipoles and adatom diffusion have been suggested as the possible mechanisms for localized field fluctuations, their predict different frequency dependence predictions \cite{24}.

The aim of this work is to study the influence of the electrostatic effect due to patch potentials between closely spaced surfaces. 1). For the spatial variations, the SAFs are typically related to the effective patch size and the variance of patch potentials over the surface. The quasi-local correlation model based on the polycrystalline surface assumption is more suitable to describe random surface potentials than others in terms of physical explanations. The SAF can be denoted by the probability that the two points are in the same grain. However, the current result of this model does not give a simple analytical expression for the form of SAF, which leads to the mechanism how the patch force is affected by patch size and distance scaling is not clear yet. These unclear points motivate us to revisit this model from a statistical perspective \cite{39}. Based on a theoretical analysis and Monte Carlo simulation, we give a cleaner relationship between the patch force and the patch size. A finite element analysis is performed to study the finite size effects of the plates, and also some assumptions used. Furthermore, a torsion balance experiment at ${\rm \mu m}$ range is conducted. The fit result shows that, a 0.4 mm effective patch size is obtained and is much larger than the empirical values (typically in the range 10 nm to 1 ${\rm \mu m}$). 2). For the temporal variations, the analysis is lacking in previous theoretical modeling. We thus introduce the time term, and analyze the fluctuation term by assuming the temporal and spatial variations are decoupled. We then apply the adatom diffusion model to describe the potential fluctuation, and compare the theoretical frequency dependence of mean potential with the experimental results provided by HUST group. Finally, we apply our model to analyze the patch effect for a typical ISL experiment at the Submillimeter range \cite{15} and LISA \cite{25}.

This paper is organized as follows. In Sec. II, we reproduce the electrostatic patch effect between two infinite parallel plates by using the Green's function method, and introduce the assumptions of patch potentials. Based on the expression we obtain, in Sec. III, we give a complete analysis for the correlation functions, including the theoretical modeling of the quasi-local models and a Monte-Carlo simulation based on Voronoi nuclei. A simple finite element analysis of the patch force is also introduced here. In Sec. IV, based on the adatom diffusion model, a temporal analysis of mean potential is presented. In Sec. V, a comparison between the theoretical model and experimental results is given. In Sec. VI, we estimate the possible influence of patch effect for two typical experiments. Our final remarks are included in Sec. VII.
\section{Basic theory for patch potential}\label{section2}
For a clean and regular surface the otherwise homogeneous density of the electrons inside the metal is distorted at the surface, which creates an effective dipole layer at the metal-air interface. This dipole layer changes the work function of surface and thus is related to the surface potential. In the limit when the layer is adjacent to the surface, this moment distribution generate variations in the potentials. Then the patch potential ${V(x, y, t)}$ of this surface can be written as \cite{11,40}
\begin{equation}\label{N1}
V(x,y,t)=P(x,y,t)/{{\varepsilon }_{0}},
\end{equation}
where ${P(x, y, t)}$ is the dipole moment density and ${{\varepsilon }_{0}}$ is the vacuum permittivity. Consider two infinite parallel plates separated by a distance ${a}$ in the ${z}$ direction. These plates are labeled as ${A}$ and ${B}$, respectively. ${\vec{r}_1}$ and ${\vec{r}_2}$ are the two-dimensional coordinates in the plates of ${A}$ and ${B}$, respectively. The common area of plates ${A}$ and ${B}$ is ${S}$. The membranes of dipole moment are located at ${z_A=h}$ and ${z_B=a-h}$, where ${2h}$ is the separation between the charges that comprise the dipole layer. The electrostatic energy in the region limited by plates ${A}$ and ${B}$ at time ${t}$ is given by
\begin{equation}\label{N2}
W(t)=\frac{1}{2}\int{\phi (\vec{r}_1,z_1, t)\rho (\vec{r_1},z_1,t)d\vec{r}_1dz_1},
\end{equation}
where ${\rho (\vec{r}_1,z,t)}$ and ${\phi (\vec{r}_1,z,t)}$ are the charge density and the electrostatic potential in this region, respectively. ${\phi (\vec{r}_1,z,t)}$ is given by
\begin{equation}\label{N3}
\phi (\vec{r}_1,z_1,t)=\frac{1}{{{\varepsilon }_{0}}}\int{\rho ({\vec{r}_2},{z_2},t)G(\vec{r}_1,z_1,{\vec{r}_2},{z_2})d{\vec{r}_2}d{z_2}},
\end{equation}
where ${G(\vec{r}_1,z_1,{\vec{r}_2},{z_2})}$ is Dirichlet Green's function for the parallel plate configuration. Inserting Eq. (\ref{N3}) into Eq. (\ref{N2}), we can obtain
\begin{equation}\label{N4}
W(t)=\frac{1}{2{{\varepsilon }_{0}}}\int{\int{\rho ({\vec{r}_2},{z_2},t)\rho (\vec{r}_1,z_1,t)G(\vec{r}_1,z_1,{\vec{r}_2},{z_2})d{\vec{r}_2}d{z_2}d\vec{r}_1dz_1}},
\end{equation}

Since the dipole moment layer is the only source in this region, the charge density can be written as follows
\begin{equation}\label{N5}
\rho (\vec{r}_1,z_1,t)={{P}_{A}}(\vec{r}_1,t){\delta }'(z_1-h)+{{P}_{B}}(\vec{r}_1,t){\delta }'(z_1-a+h),
\end{equation}
where ${{\delta }'(z_1)}$ is the first derivative of Dirac's function. Using the relationship in Eq. (\ref{N3}), the energy is
\begin{equation}\label{N6}
W(t)=\frac{{{\varepsilon }_{0}}}{2}\sum\limits_{A=1}^{2}{\sum\limits_{B=1}^{2}{\int{\int{d{\vec{r}_2}d{z}'d\vec{r}_1dz_1{{V}_{A}}(\vec{r}_1,t){{V}_{B}}({\vec{r}_2},t){{\left. \frac{{{\partial }^{2}}G(\vec{r}_1,z,{\vec{r}_2},{z_2})}{\partial z_1\partial {z_2}} \right|}_{z_1={{z}_{A}}+h,{z_2}={{z}_{B}}+h}}}}}},
\end{equation}
where ${{{V}_{A}}(\vec{r}_1,t)}$ and ${{{V}_{B}}(\vec{r}_2,t)}$ are the observable potential of plates ${A}$ and ${B}$, respectively. Now the target is to obtain the formula of ${G(\vec{r}_1,z_1,{\vec{r}_2},{z_2})}$. Using the method in Ref. \cite{41}, ${G(\vec{r}_1,z_1,{\vec{r}_2},{z_2})}$ is easily computed by using Fourier transform method
\begin{equation}\label{N7}
G(\vec{r},z_1,{\vec{r}_2},{z_2})={{G}_{0}}(\vec{r}_1,z_1,{\vec{r}_2},{z_2})+\frac{1}{8{{\pi }^{2}}}\int{\frac{{{e}^{j\vec{k}(\vec{r}_1-{\vec{r}_2})}}}{k\sinh ka}[\sinh k(a-{z_2})\cdot {{e}^{-kz_1}}+\sinh k{z_2}\cdot {{e}^{-k(a-z_1)}}]d\vec{k}},
\end{equation}
where ${{{G}_{0}}(\vec{r}_1,z_1,{\vec{r}_2},{z_2})}$ is the Green's function in the absence of boundaries, ${\vec{k}}$ is the corresponding variable of ${\vec{r} = \vec{r}_1-{\vec{r}_2}}$ in Fourier spaces and ${k=\left| {\vec{k}} \right|}$. In this limit ${h\to 0}$, we finally get
\begin{equation}\label{N8}
W(t)=\frac{{{\varepsilon }_{0} S}}{8{{\pi }^{2}}}\int{\frac{[{{C}_{AA}}(\vec{k},t)+{{C}_{BB}}(\vec{k},t)]\cosh ka-{{C}_{AB}}(\vec{k})-{{C}_{BA}}(\vec{k})}{\sinh ka}}kd\vec{k},
\end{equation}
where ${{{C}_{AB}}(\vec{k},t)=\int{{{C}_{AB}}(\vec{r},t){{e}^{j\vec{k} \vec{r}}}d\vec{r}}=\frac{1}{S}\int{\int{{{V}_{A}}({\vec{r}_2}+\vec{r},t){{V}_{B}}({\vec{r}_2},t){{e}^{j\vec{k} \vec{r}}}d{\vec{r}_2}d\vec{r}}}}$ can be regarded as the two-dimensional Fourier transform of the correlation function of the potential of plates ${A}$ and ${B}$. Then the force along axis z can be given by
\begin{equation}\label{N9}
{{F}_{z}}(t)=\frac{{{\varepsilon }_{0} S}}{8{{\pi }^{2}}}\int{\frac{{{C}_{AA}}(\vec{k},t)+{{C}_{BB}}(\vec{k},t)-[{{C}_{AB}}(\vec{k},t)+{{C}_{BA}}(\vec{k},t)]\cosh ka}{{{\sinh }^{2}}ka}}{{k}^{2}}d\vec{k},
\end{equation}

This formula is consistent with previous results in the literatures except for the time term \cite{11, 12}. One can recover the usual result for perfect conductors by assuming that ${{{V}_{A}}(\vec{r}_1,t) = {{\bar{V}}_{A}}(t)}$ and ${{{V}_{B}}(\vec{r}_2,t) = {{\bar{V}}_{B}}(t)}$. In this case,  ${{{C}_{AB}}(\vec{k},t)=4{{\pi }^{2}}{{{\bar{V}}_{A}}(t)}{{\bar{V}}_{B}(t)}{{\delta }^{2}}(\vec{k})}$, which leads to
\begin{equation}\label{N10}
{{F}_{z}}(t)=\frac{{{\varepsilon }_{0}}S}{2{{a}^{2}}}{{({{\bar{V}}_{A}(t)}-{{\bar{V}}_{B}(t)})}^{2}}.
\end{equation}
where ${{{{\bar{V}}_{A}(t)}-{{\bar{V}}_{B}(t)}}}$ is referred to as contact potential differences (CPD). In this limit, the model reduces to the parallel-plate capacitor model. This long range force is relatively easily calculated, and the CPD can be eliminated by applying compensation voltage in real experiment. Please noted that the CPD is a function of time ${t}$, which means that a time monitor of voltage is needed for every data run. Therefore, it is convenient to assume that the potentials of plates have only stochastic components fluctuating around 0, i.e., ${{{\left\langle {{V}_{A/B}}(\vec{r}_1,t) \right\rangle }_{S}}=0}$ \cite{18}. The purpose of this paper is to study the influence of the random patch force.

In order to evaluate Eq. (\ref{N9}), we have to make some assumptions about the nature of the patch potentials. One assumption is that the spatial and temporal variations of the potentials decouple. Another one is that the surface can be divided into ${N}$ separate patches based on the potential difference. Providing that ${N}$ is large and the stochastic process is stationary and ergodic, we can write the patch potential over surface ${A}$ as \cite{23}
\begin{equation}\label{N11}
{{V}_{A}}(\vec{r}_1,t)=\sum\limits_{i=1}^{N}{{{v}_{A,i}}(t){{\chi }_{i}}(\vec{r}_1)},
\end{equation}
where ${{{v}_{A,i}}(t)}$ is the fluctuating potential of ${i}$th patch, and the step function ${{{\chi }_{i}}(\vec{r}_1)}$ is 1 only for ${\vec{r}_1}$ within the area of the ${i}$th patch, and 0 otherwise. Then the correlation function of the potentials in plate ${A}$ can be further expressed as
\begin{equation}\label{N12}
{{C}_{AA}}(\vec{r},t)=\frac{1}{S}\int{\sum\limits_{i=1}^{N}{{{v}_{A,i}}(t){{\chi }_{i}}({{{\vec{r}}}_{1}}+\vec{r})}\sum\limits_{j=1}^{N}{{{v}_{A,j}}(t){{\chi }_{j}}({{{\vec{r}}}_{1}})}d{{{\vec{r}}}_{1}}},
\end{equation}

This integration can be divided into ${N}$ patches. Each patch ${i}$ has area ${S_i}$. In this case, it becomes
\begin{equation}\label{N13}
{{C}_{AA}}(\vec{r},t)\simeq\frac{\sum\limits_{k=1}^{N}{\int_{S_k}{{v_{A,k}}(t)\sum\limits_{i=1}^{N}{{v}_{A,i}}(t){{\chi}_i}({\vec{r}_{k}}+\vec{r})d{{\vec{r}}_k}}}}{S},
\end{equation}

As usual, the variance of patch potentials over the surface ${A}$ can be given by setting ${\vec{r} = (0,0)}$
\begin{equation}\label{N14}
V_{rms}^{2}(t)={{C}_{AA}}(0,t)\simeq \frac{\sum\limits_{k=1}^{N}{\int_{{{S}_{k}}}{v_{A,k}^{2}(t)d{{{\vec{r}}}_{k}}}}}{S},
\end{equation}

Now recalling Eq. (\ref{N13}), the integral term can be rewritten as
\begin{equation}\label{N15}
\int_{{{S}_{k}}}{{{v}_{A,k}}(t)\sum\limits_{i=1}^{N}{{{v}_{A,i}}}(t){{\chi }_{i}}({{{\vec{r}}}_{k}}+\vec{r})d{{{\vec{r}}}_{k}}}=\int_{{{S}_{k}}}{v_{A,k}^{2}(t){{\chi }_{k}}({{{\vec{r}}}_{k}}+\vec{r})d{{{\vec{r}}}_{k}}}+{{\left. \int_{{{S}_{k}}}{{{v}_{A,k}}(t){{v}_{A,i}}(t){{\chi }_{i}}({{{\vec{r}}}_{k}}+\vec{r})d{{{\vec{r}}}_{k}}} \right|}_{i\ne k}},
\end{equation}

Physically, the potential of each patch is statistically independent. Eq. (13) can be further expressed as
\begin{equation}\label{N16}
{{C}_{AA}}(\vec{r},t)\simeq \frac{\sum\limits_{k=1}^{N}{\int_{{{S}_{k}}}{v_{A,k}^{2}(t){{\chi }_{k}}({{{\vec{r}}}_{k}}+\vec{r})d{{{\vec{r}}}_{k}}}}}{S}\simeq V_{rms}^{2}(t)W(\vec{r}).
\end{equation}
where ${W(\vec{r})}$ is the probability that points  ${{\vec{r}}_{1}}$ and ${{\vec{r}}_{2}}$ are in the same patch with ${\vec{r}={{\vec{r}}_{1}}-{{\vec{r}}_{2}}}$. Noticed that, under the assumption of Eq. (\ref{N11}), Eq. (\ref{N16}) is almost identical to the result in Ref. \cite{19} except for the ${t}$ term. It tell us that a repeated potential measurement is needed even at the same place of plate. Now the target is to obtain the explicit form of ${W(\vec{r})}$. Before entering the next section, one can discuss a special case for ${W(\vec{r})}$. If we assume that the size of the patch is so small, i.e., the point patch. In this case, ${W(\vec{r})=\frac{S}{N}{{\delta }^{2}}(\vec{r})}$ and ${{{C}_{AB}}(\vec{k},t)=\frac{S}{N}V_{rms}^{2}(t)}$. Usually, there are no cross correlations between the patches on different plates ${A}$ and ${B}$ have similar patch distribution. Then we have
\begin{equation}\label{N17}
{{F}_{z,pp}}(t)=\frac{{{\varepsilon }_{0}}S}{4{{\pi }^{2}}}\frac{S}{N}V_{rms}^{2}(t)\int{\frac{{{k}^{2}}}{{{\sinh }^{2}}ka}}d\vec{k}\approx 0.90\frac{{{\varepsilon }_{0}}S}{\pi }\frac{S}{N}\frac{V_{rms}^{2}(t)}{{{a}^{4}}}.
\end{equation}
Although this ${1/a^4}$ scaling law shows a stronger relationship with distance, the magnitude of force is suppressed by the number of patches. Therefore, we need more information about the relationship between the force and the patch size.

\section{Spatial variation of the surface potential}\label{section3}
\subsection{theoretical modeling of the random electrostatic force correlation function}
The analytical way to calculate ${W(\vec{r})}$ is employing expressions for the probability density of the effective patch length. As mentioned before, the patch correlation function has been studied by Behunin ${et \, al}$ \cite{19}. They used a quasi-local correlation model with a better physical explanation than the sharp-off model to describe this phenomenon in Casimir experiment. However, they did not give a simple analytical expression for the final form of ${W(r)}$, nor did ${W(k)}$, which leads the relationship between the patch size and the distance to be ambiguous. Motivated by this, we revisit this model from a statistical perspective and find some unnoticed points. Following the discussion in Sec. II, the surface is divided into ${N}$ separate patches, ${N}$ is large and the stochastic process is stationary and ergodic. To obtain an analytical expression of ${W(\vec{r})}$, we denote the random variable ${\xi }$ by two possible outputs
\begin{equation}\label{N18}
\xi ({{\vec{r}}_{1}},\vec{r})=\left\{ \begin{matrix}
   1, \; O1  \\
   0, \; O2  \\
\end{matrix} \right.,
\end{equation}
where ${O1}$ is defined to be the output for the points 1 and 2 are in the same patch, and ${O2}$ otherwise. Therefore, ${W(\vec{r})}$ can be regarded as the spatial average of ${\xi }$ by assuming the stochastic process is stationary
\begin{equation}\label{N19}
W(\vec{r})={{\left\langle \xi ({{{\vec{r}}}_{1}},\vec{r}) \right\rangle }_{S}}=\underset{S\to \infty }{\mathop{\lim }}\,\frac{1}{S}\int_{S}{\xi ({{{\vec{r}}}_{1}},\vec{r})d{{{\vec{r}}}_{1}}},
\end{equation}	

This integration can be divided into ${N}$ patches. Each patch ${i}$ has area ${S_i}$. In this case, it becomes
\begin{equation}\label{N20}
W(\vec{r})=\underset{N\to \infty }{\mathop{\lim }}\,\frac{\frac{1}{N}\sum\limits_{i=1}^{N}{\int_{{{S}_{i}}}{\xi ({{{\vec{r}}}_{i}},\vec{r})d{{{\vec{r}}}_{i}}}}}{\frac{1}{N}\sum\limits_{i=1}^{N}{{{S}_{i}}}},
\end{equation}	

By assuming a probability density function ${p(s)}$, Eq. (\ref{N20}) can be written as
\begin{equation}\label{N21}
W(\vec{r})=\frac{\int{p(s){{d}^{2}}s\int_{{{S}_{i}}}{\xi ({{{\vec{r}}}_{1}},\vec{r})d{{{\vec{r}}}_{1}}}}}{\int{s}p(s){{d}^{2}}s},
\end{equation}	
From Eq. (\ref{N21}), we can know that ${W(\vec{r})}$ is depending on the shape of patch. We can assume that the patches are isotropic, i.e., that autocorrelations are spherically symmetric. This assumption leads to two simplifying situations. One is that we only choose the direction of displacement ${r}$ along one direction, i.e., ${r = x}$ or ${r = y}$ \cite{39}. It means that every single patch can be approximated as one line with cord length ${l}$. Here we refer it as one-dimensional quasi-local model. Eq. (\ref{N21}) can be rewritten as
\begin{equation}\label{N22}
{{W}_{1}}(r)=\frac{\int{\int{\xi ({{x}_{1}},r)p(l)d{{x}_{1}}}}dl}{\int{l}p(l)dl},
\end{equation}	

In this case, ${\xi ({{x}_{1}},r)=\theta (l/2-{{x}_{1}})\theta (l/2-r+{{x}_{1}})}$ and ${\theta ({{x}_{1}})}$ is the unit step function. Then we have
\begin{equation}\label{N23}
{{W}_{1}}(r)=\frac{\int_{r}^{\infty }{(l-r)p(l)dl}}{\int{l}p(l)dl},
\end{equation}

Usually, we assume that the cord length ${l}$ has Poisson statistic, which is a good approximation due to the "memoryless" property \cite{34}. The resulting density function is
\begin{equation}\label{N24}
p(l)={{e}^{-l/\lambda }}/\lambda,
\end{equation}
where ${\lambda}$ is the constant of proportionality. Inserting Eq. (\ref{N24}) into (\ref{N23}), we can obtain
\begin{equation}\label{N25}
{{W}_{1}}(r)={{e}^{-r/\lambda }},
\end{equation}
\begin{figure*}[htbp]
\centering
\subfigure{%
    \includegraphics[width=3.1in]{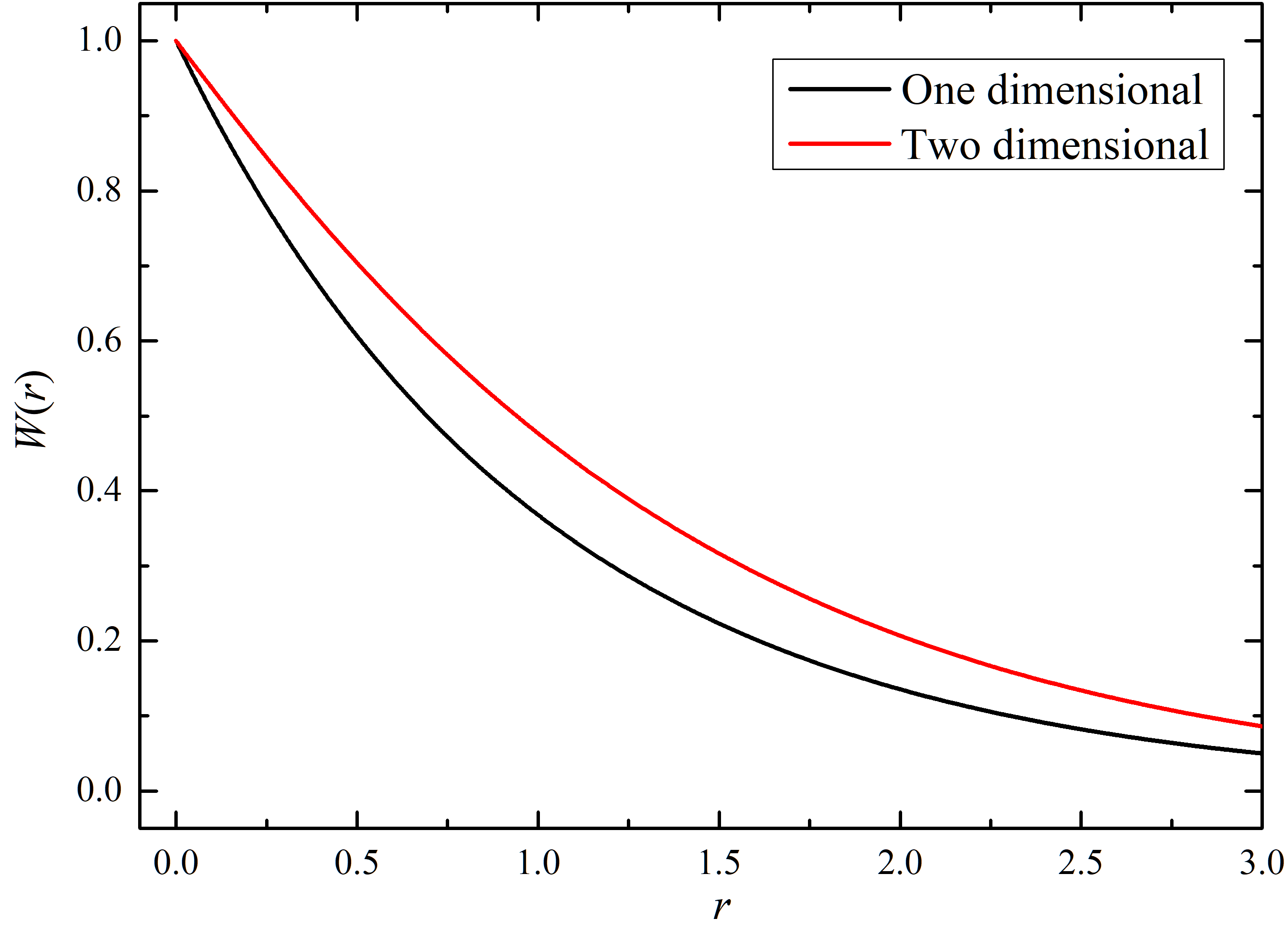}}
\quad
\subfigure{
    \includegraphics[width=3.1in]{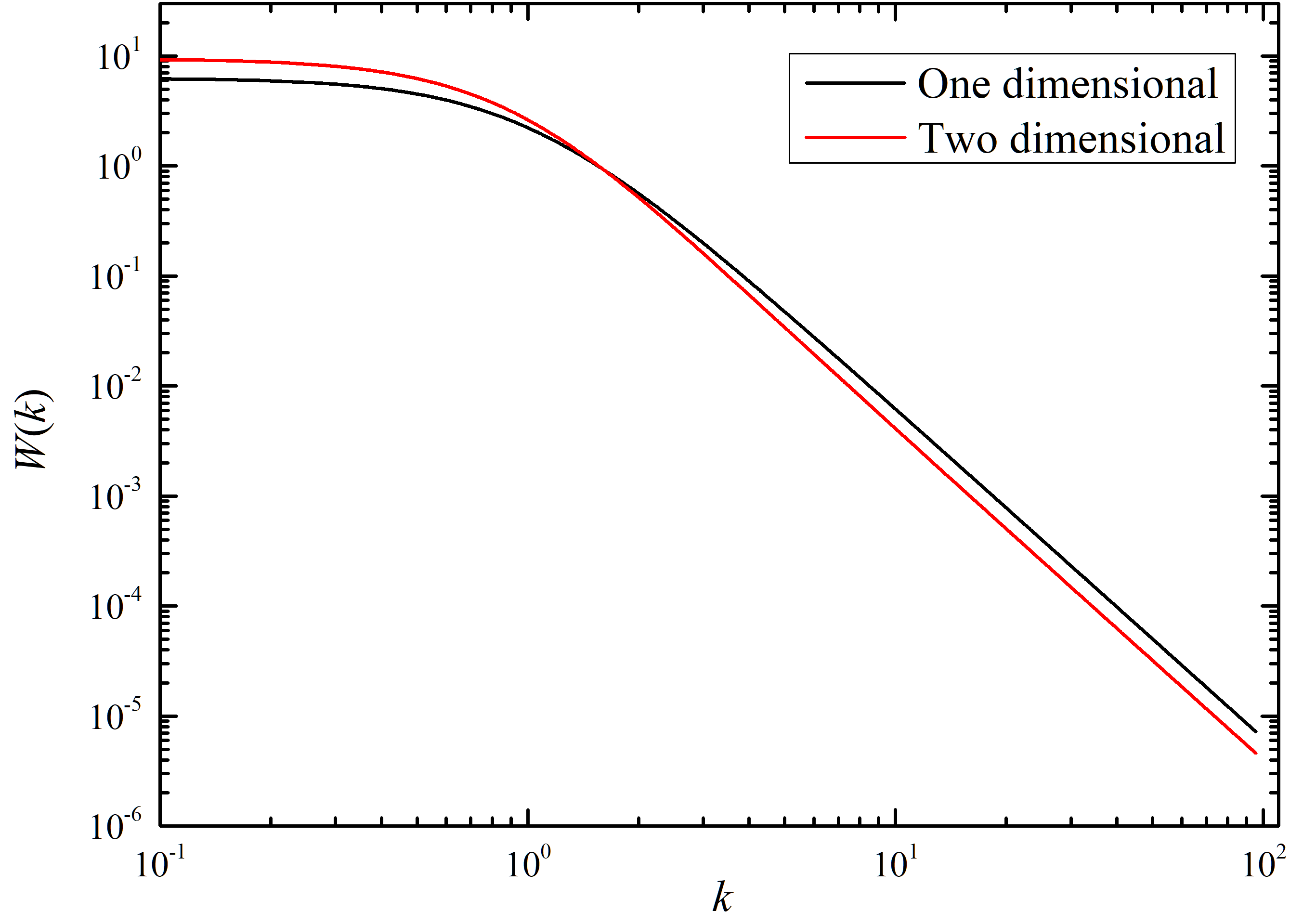}}
\caption{Comparison of the one-dimensional and two-dimensional model in real space and Fourier space with ${\lambda = 1}$.}
\label{fig1}
\end{figure*}
This is the result that have been used in many literatures, such as Refs. \cite{13,18,19,21}. The advantage of this model is that a simple form of ${W(k)}$ in Fourier space can be obtained. By using the zero-order Hankel transforms, we have
\begin{equation}\label{N26}
{{W}_{1}}(k)=\frac{2\pi {{\lambda }^{2}}}{{{(1+{{k}^{2}}{{\lambda }^{2}})}^{3/2}}}.
\end{equation}

Another situation is that the patch is circular with a radius ${l}$. We refer it as two-dimensional quasi-local model. It is worth emphasizing that the circular cannot fill surface, the gaps have been ignored in this analysis. Eq. (\ref{N21}) can be rewritten as
\begin{equation}\label{N27}
{{W}_{2}}(r)=\frac{4}{\pi }\frac{\int{\int_{{{S}_{i}}}{\xi ({{{\vec{r}}}_{1}},\vec{r})d{{{\vec{r}}}_{1}}}}p(l)dl}{{{{\bar{l}}}^{2}}},
\end{equation}
where ${\xi ({{\vec{r}}_{1}},\vec{r})=\theta (l/2-\left| {{{\vec{r}}}_{1}} \right|)\theta (l/2-\left| \vec{r}-{{{\vec{r}}}_{1}} \right|)}$. Thus
\begin{equation}\label{N28}
{{W}_{2}}(r)=\frac{2}{\pi {{{\bar{l^{2}}}}}}\int_{r}^{\infty }{\left[ {{l}^{2}}\arccos (\frac{r}{l})-r\sqrt{{{l}^{2}}-{{r}^{2}}} \right]p(l)dl},
\end{equation}
It should note that ${\bar{l^{2}}}$ is used in the denominator, rather than ${l^{2}}$ in Ref. \cite{19}. Similarity, if a Poisson statistic of ${l}$ is used, we have
\begin{equation}\label{N29}
{{W}_{2}}(r)=\frac{1}{\pi }\left[ 2G_{2,4}^{4,0}(\frac{{{r}^{2}}}{4{{\lambda }^{2}}})-\frac{{{r}^{2}}}{{{\lambda }^{2}}}{{K}_{1}}(\frac{r}{\lambda }) \right],
\end{equation}
where ${G_{p,q}^{m,n}(x)}$ is the Meijer's G-Function and ${{{K}_{1}}(x)}$ is the first-order modified Bessel function \cite{42}. We also can obtain the Fourier transform of
\begin{equation}\label{N30}
{{W}_{2}}(k)=4\frac{(1+{{k}^{2}}{{\lambda }^{2}})EpK(-{{k}^{2}}{{\lambda }^{2}})-(1-{{k}^{2}}{{\lambda }^{2}})EpE(-{{k}^{2}}{{\lambda }^{2}})}{{{(k+{{k}^{3}}{{\lambda }^{2}})}^{2}}}.
\end{equation}
where ${EpK(x)}$ and ${EpE(x)}$ are the complete elliptic integral of the first kind and of the second kind, respectively. One can make some check for Eqs. (\ref{N25}) and (\ref{N30}) by using ${\int_{0}^{\infty }{kW(k)dk}=2\pi}$. Note that, one-dimensional situation has a simple correlation function than two-dimensional quasi-local model, but two-dimensional quasi-local model has a stronger physical explanation. FIG. \ref{fig1} shows the comparison between the one-dimensional and two-dimensional quasi-local model with ${\lambda = 1}$. We can see that the difference exists and the two-dimensional model corresponds to a stronger correlation.

\begin{figure*}[htbp]
\centering
\subfigure{%
    \includegraphics[width=3.05in]{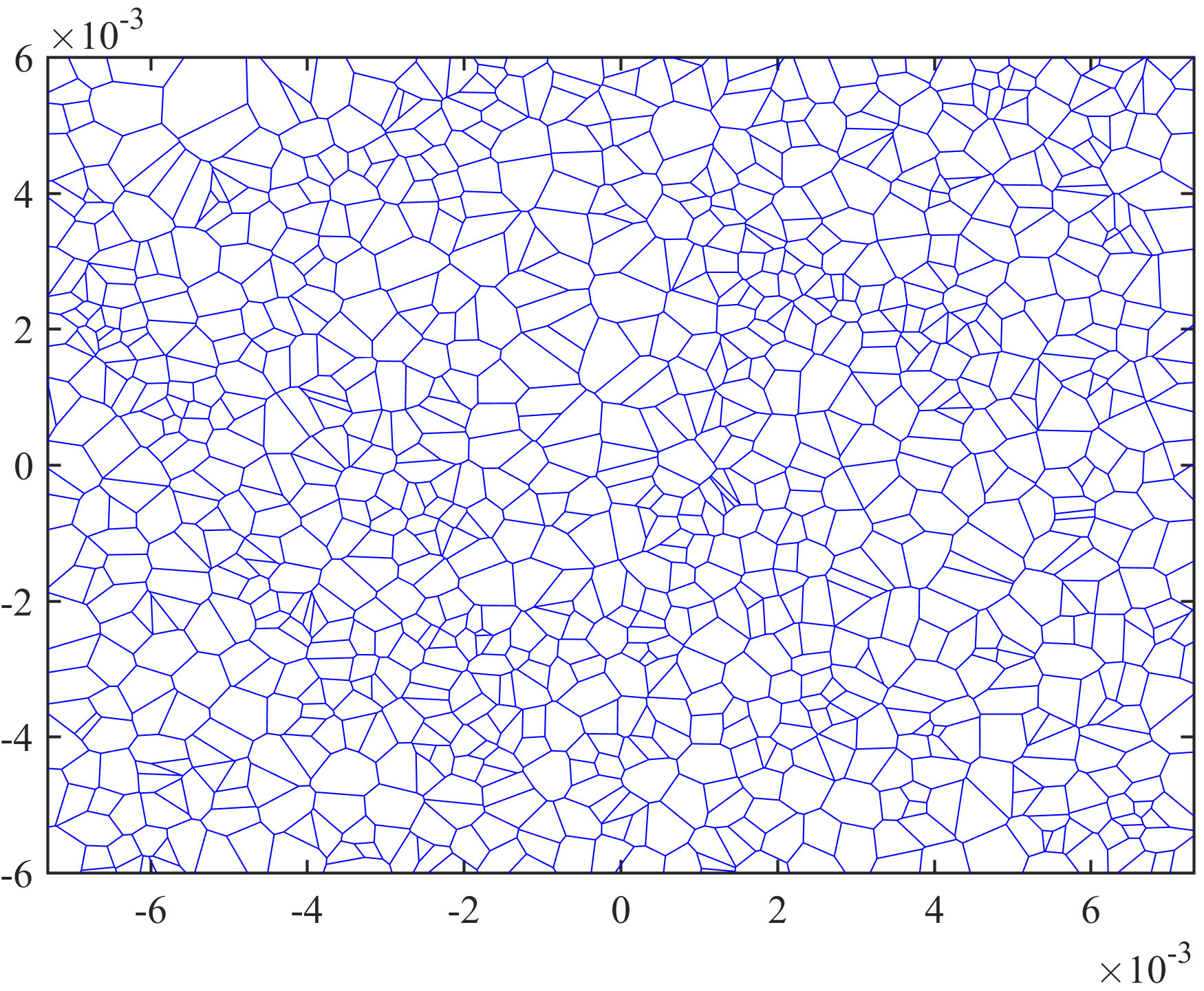}}
\quad
\subfigure{
    \includegraphics[width=3.15in]{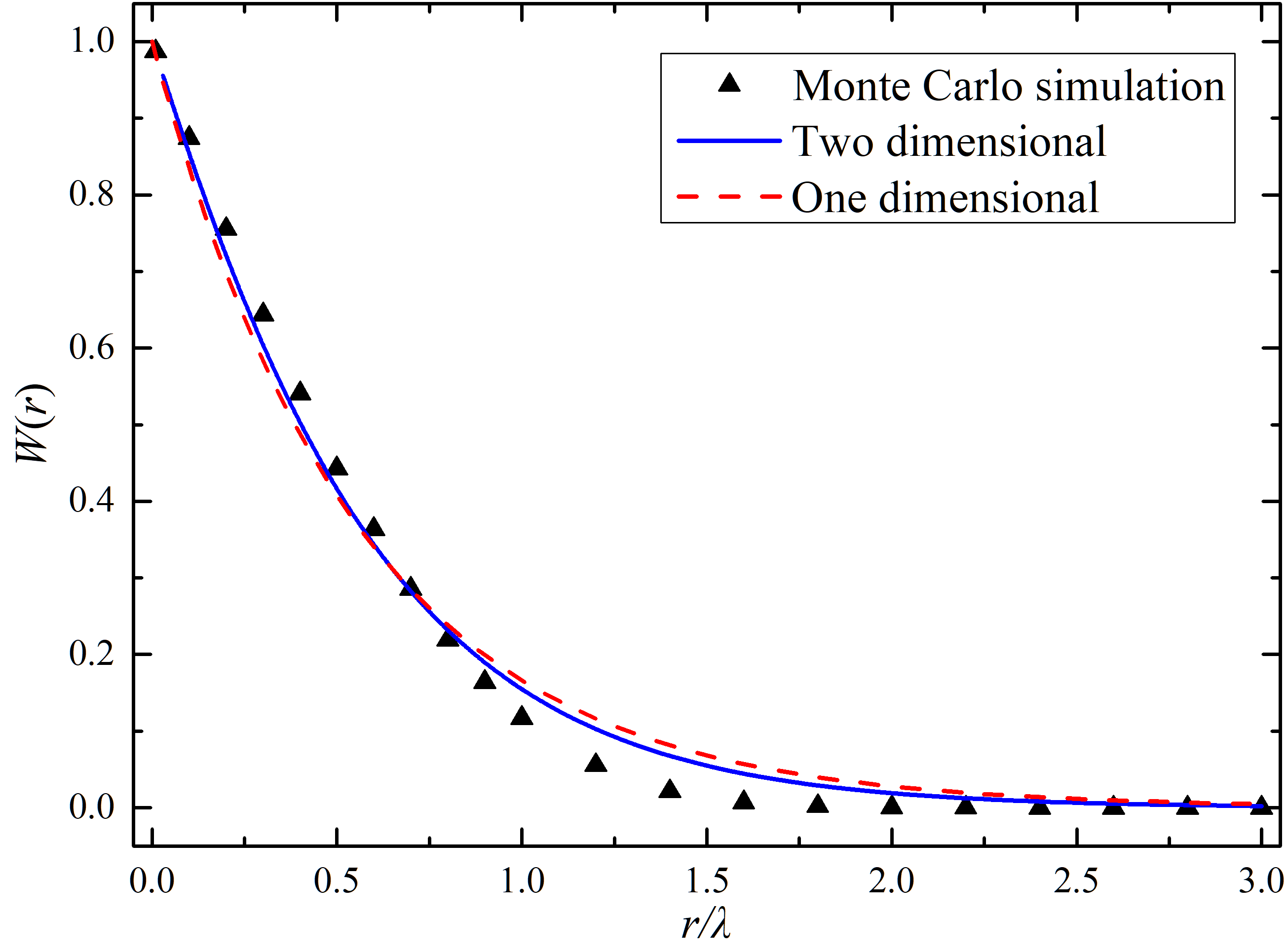}}

\caption{(left) 2-D diagrams of Voronoi polycrystals. (right) Comparison of two different least-squares-fit models with the simulation result.}
\label{fig2}
\end{figure*}

In order to verify the correctness of the above assumptions, we performed a Monto Carlo simulation to obtain the specific values of ${W(r)}$. This simulation is relatively straightforward based on Eq. (\ref{N20}). Firstly, we construct ${N}$ Voronoi nucleus in a fixed surface with area ${S}$ by using the method introduced by Debye, as shown in FIG. \ref{fig2} (left). The basic principle of constructing Voronoi nuclei is determining the area with the closet center point. After finishing this procedure, we select point ${r_1}$ randomly within the surface. Then we select another point ${r_2}$ with a fixed distance ${r}$ from ${r_1}$. Finally, we determine whether both points of each pair lie within a single grain. To make the results convergence to below ${1 \%}$, the procedures are repeated ${10^5}$ times for every value of distance. The distance ${r}$ is incrementally varied to produce a discrete sampling of ${W(r)}$. In addition, we use ${\lambda \to w/q}$ to obtain a more obvious statistical property since we have no information about the value of ${N}$, where ${w=\sqrt{S/N}}$ and ${q}$ is a dimensionless constant to be determined. Therefore, the simulation result can be expressed as a function of ${r/w}$, as shown in FIG. \ref{fig2} (right) (triangle shape). We use Eqs. (\ref{N25}) and (\ref{N28}) to fit the result and find that the covariances are good (red dashed line and blue solid line). The best least-squares fits give ${q = 1.79}$ and ${q = 2.33}$, respectively. Although the fit results of these two models are similar, the one-dimensional quasi-local model correspond to a bigger patch size. Therefore, we adopt Eqs. (\ref{N29}) and (\ref{N30}) in the sequel calculations. In addition, we also find that the values of ${q}$ are almost identical for different ${N}$. This result proves that the procedure of replacing ${\lambda}$ with ${w/q}$ is necessary.

We now employ the two-dimensional quasi-local model to obtain the properties of the random patch force. We will also assume that there are no cross correlations between the patches on different plates ${A}$ and ${B}$ have similar patch distribution. Inserting Eqs. (\ref{N16}) and (\ref{N30}) into Eq. (\ref{N9}), which leads to
\begin{equation}\label{N31}
{{F}_{z}}=\frac{2{{\varepsilon }_{0}}}{\pi }\frac{V_{rms}^{2}S}{{{a}^{2}}}\int_{0}^{\infty }{\frac{{{k}^{3}}}{{{\sinh }^{2}}k}}\Omega (k,\frac{a}{w})dk,
\end{equation}
where
\begin{equation}\label{N32}
\Omega (k,\frac{a}{w})=\frac{(1+{{k}^{2}}\frac{1}{{{q}^{2}}}\frac{{{w}^{2}}}{{{a}^{2}}})EpK(-{{k}^{2}}\frac{1}{{{q}^{2}}}\frac{{{w}^{2}}}{{{a}^{2}}})-(1-{{k}^{2}}\frac{1}{{{q}^{2}}}\frac{{{w}^{2}}}{{{a}^{2}}})EpE(-{{k}^{2}}\frac{1}{{{q}^{2}}}\frac{{{w}^{2}}}{{{a}^{2}}})}{{{(k+{{k}^{3}}\frac{1}{{{q}^{2}}}\frac{{{w}^{2}}}{{{a}^{2}}})}^{2}}}.
\end{equation}
where ${q=2.33}$ and variable substitution ${k a =k}$ has been used.
\begin{figure*}[htbp]
\centering
\subfigure{%
    \includegraphics[width=3.1in]{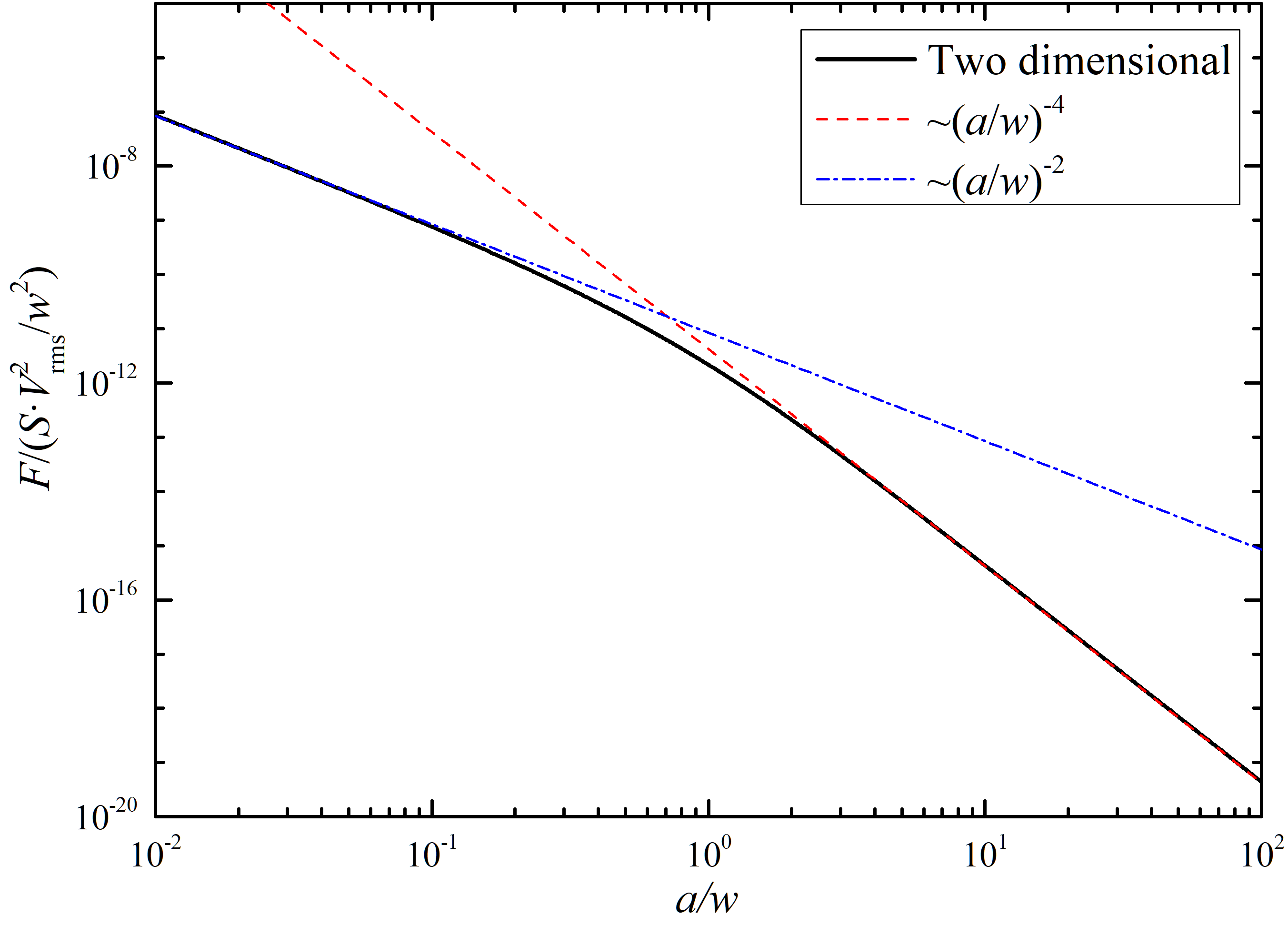}}
\quad
\subfigure{
    \includegraphics[width=3.1in]{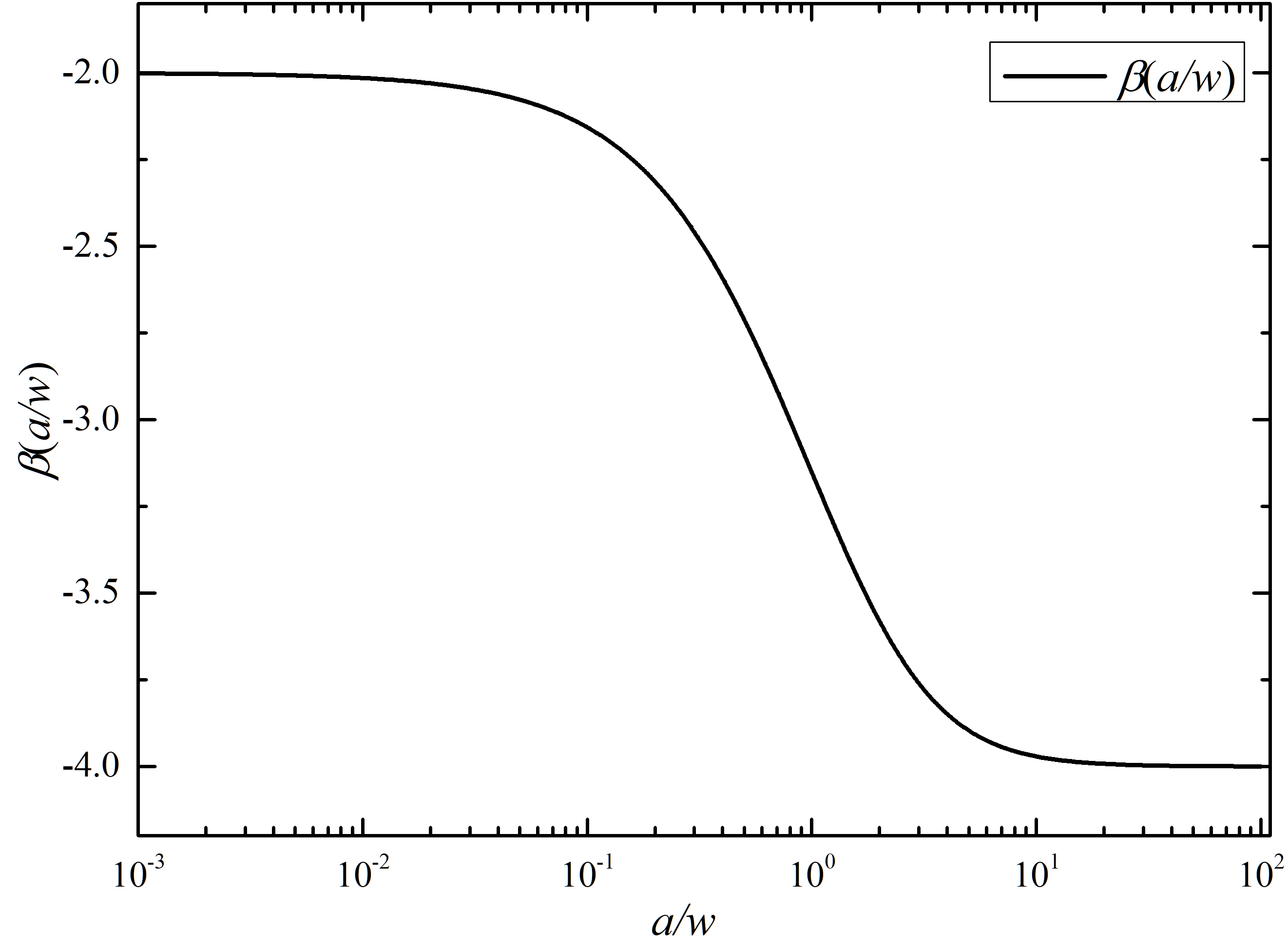}}

\caption{(left) Normalized random patch force as ${a/w}$ varies (black solid line), for ${a>>w}$ one has ${{{F}_{z}}=1.5{{\pi }^{\text{-}1}}{{\varepsilon }_{0}}V_{rms}^{2}S{{w}^{2}}{{d}^{-4}}}$ (red dashed line), and for ${a<<w}$ one has ${{{F}_{z}}=3{{\pi }^{\text{-}1}}{{\varepsilon }_{0}}V_{rms}^{2}S{{d}^{-2}}}$ (blue dot-dashed line). (right) The scaling coefficient ${\beta (a/w)}$ as a function ${a/w}$.}
\label{fig3}
\end{figure*}
Therefore, we can obtain the relationship between the random patch force and ${a/w}$, as shown in FIG. \ref{fig3} (left). Furthermore, we focus on two important limits: in the case of ${a>>w}$, we obtain ${{{F}_{z}}=1.5{{\pi }^{\text{-}1}}{{\varepsilon }_{0}}V_{rms}^{2}S{{w}^{2}}{{d}^{-4}}}$ and, in the case of ${a<<w}$, we obtain ${{{F}_{z}}=3{{\pi }^{\text{-}1}}{{\varepsilon }_{0}}V_{rms}^{2}S{{d}^{-2}}}$. To obtain a simpler expression of Eq. (\ref{N31}), we introduce a scaling coefficient, such that
\begin{equation}\label{N33}
{{F}_{z}}(\frac{a}{w})=\frac{2{{\varepsilon }_{0}}SV_{\text{rms}}^{2}f(\beta )}{\pi {{a}^{2}}}{{(\frac{a}{w})}^{\beta (a/w)+2}},
\end{equation}
where ${f(\beta )}$ is a correction factor and ${\beta (a/w)}$ can be determined by
\begin{equation}\label{N34}
\beta (a/w)=\frac{\partial \ln {{F}_{z}}}{\partial \ln \left[ a/w \right]}\in (-4,-2).
\end{equation}
Eq. (\ref{N33}) is the main result of this paper. We also can plot ${\beta (a/w)}$ as a function of ${a/w}$, as shown in FIG. \ref{fig3} (right). One can obtain that once ${a/w>10}$, ${\beta =-4}$ and ${a/w<0.1}$, ${\beta =-2}$.

Similarity, for the force gradient along axis ${z}$, we have
\begin{equation}\label{N35}
{{K}_{e}}=\frac{4{{\varepsilon }_{0}}}{\pi }\frac{V_{rms}^{2}S}{{{a}^{3}}}\int_{0}^{\infty }{\frac{{{k}^{4}}\cosh k}{{{\sinh }^{3}}k}}\Omega (k,\frac{a}{w})dk,
\end{equation}
then
\begin{equation}\label{N36}
{{K}_{e}}=\frac{4{{\varepsilon }_{0}}SV_{rms}^{2}f(\beta )}{\pi {{a}^{3}}}{{(\frac{a}{w})}^{\chi (a/w)+3}},
\end{equation}
where
\begin{equation}\label{N37}
\chi (a/w)=\frac{\partial \ln {{K}_{e}}}{\partial \ln \left[ a/w \right]}\in (-5,-3).
\end{equation}

According to the Monte Carlo simulation, we give a clear relationship between the force and the patch size. One can obtain the effective patch size through the distance scaling coefficient. Generally, values of patch size reported by experiments are in the range 10 nm to 1 ${\rm \mu m}$. Therefore, we can choose suitable distance based on the empirical patch sizes. However, patch size is very dependent on material and surface preparation. An additional calibration estimation is needed for specific experiment.

\subsection{the finite element analysis of the random electrostatic force}
In the above discussion, we neglect the finite size effects of the plates. In order to make our analysis more precisely, we perform a finite element analysis (FEA) to explore this effect by using a commercial software package (COMSOL Multiphysics) \cite{33}. The package used in our simulations is AD/DC module and the CAD models are created by using the parameters in Sec. II. The distances between plates are set at ${\rm \mu m}$ level.

\begin{figure*}[htbp]
\centering
\subfigure{%
    \includegraphics[width=3.0in]{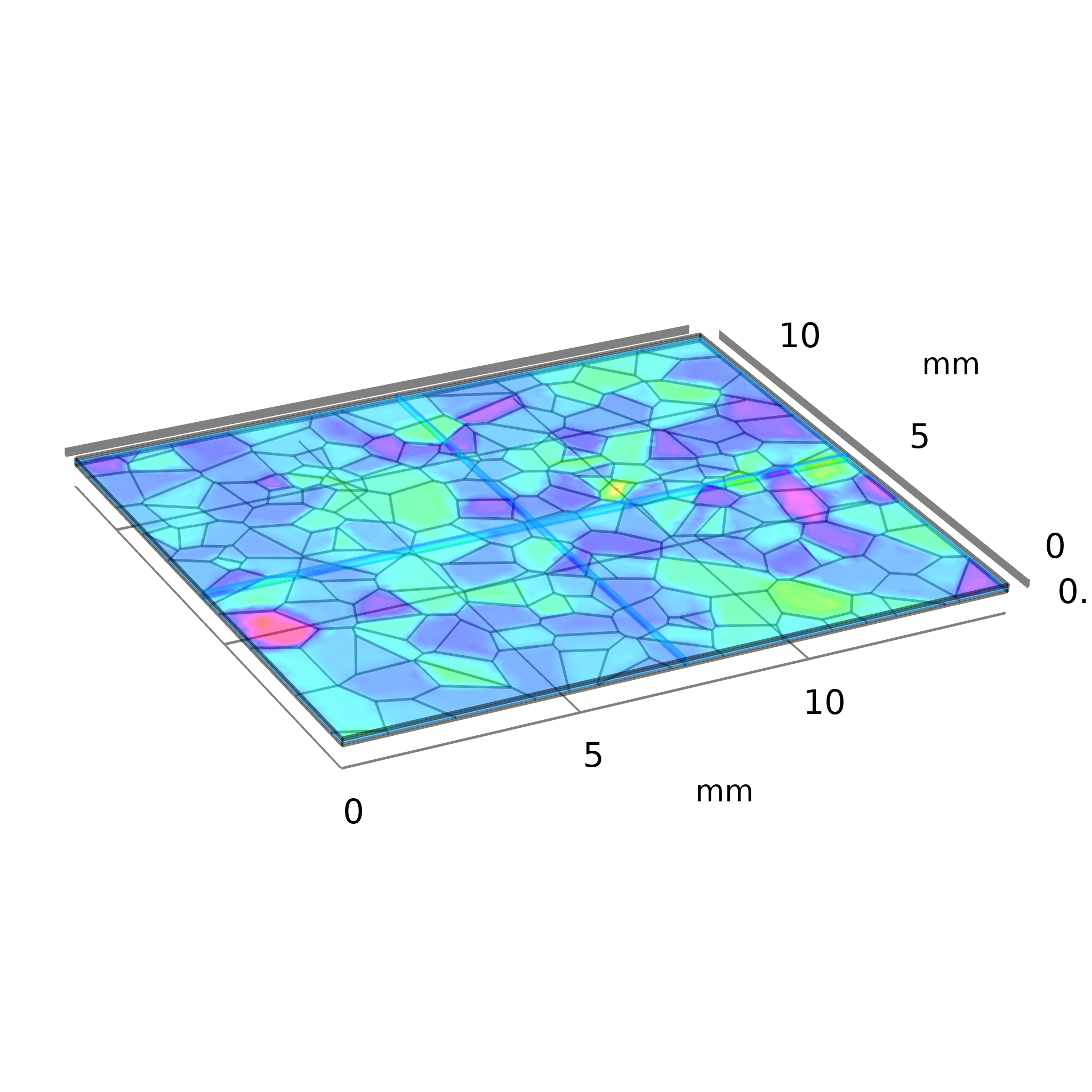}}
\quad
\subfigure{
    \includegraphics[width=3.0in]{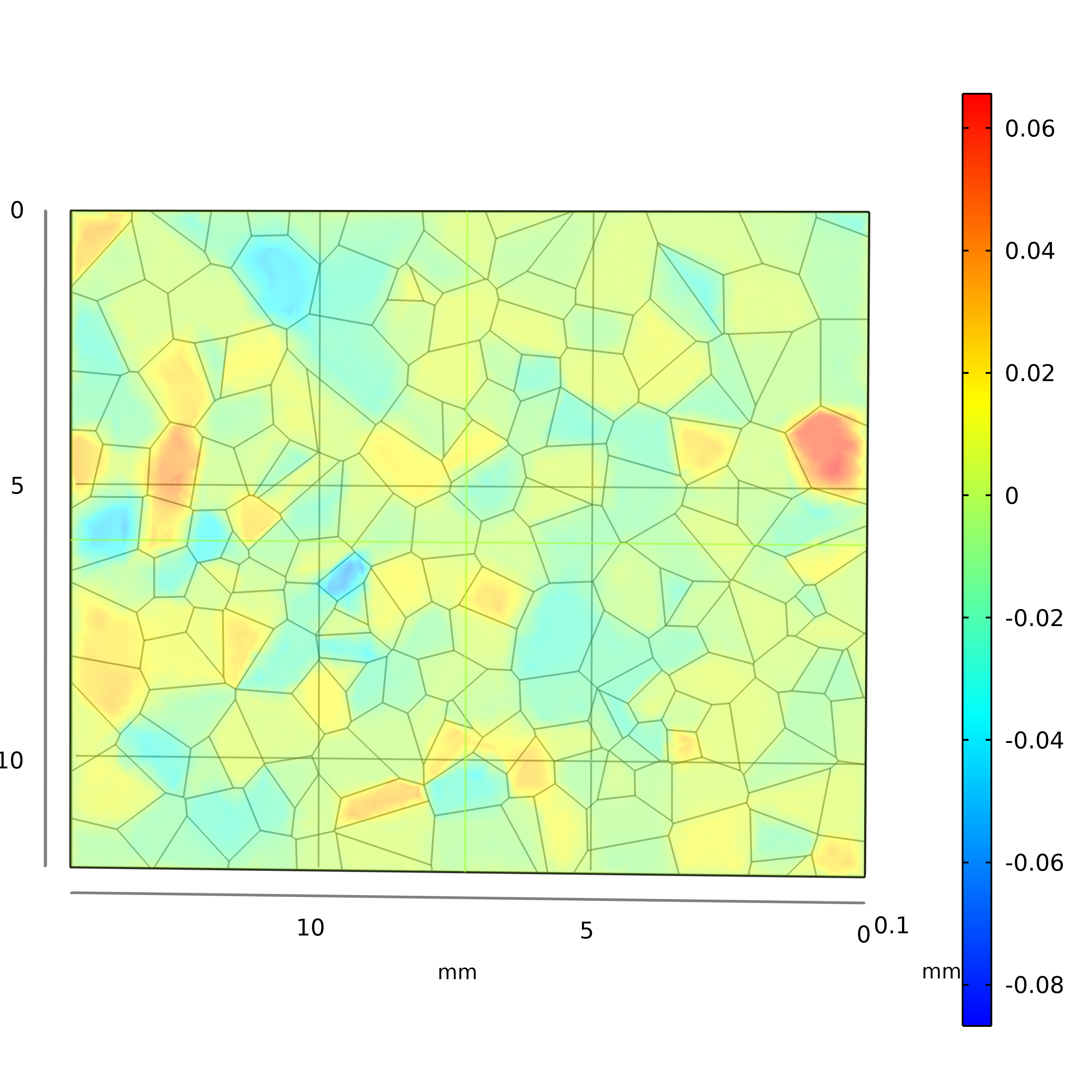}}

\caption{\label{fig4} (left) The CAD model drawn in COMSOL. (right) a Voronoi surface with 200 patches with standard deviation of voltage 30 mV.}
\end{figure*}

We first draw two rectangular boxes to represent the plates. The boxes are placed inside a big vacuum solution domain held at zero potential. Following the discussion in Sec. III, we build one surface based on the Voronoi polycrystals in COMSOL. Then this surface is placed on the opposite side of the two boxes, as shown in FIG. \ref{fig4}(left). The number of the patch changes with our needs. The potentials of the plates are set to 0. Random potentials with variance are assigned to the patches by using a random number generator. For example, we obtain a Voronoi surface with 200 patches with standard deviation of voltage 30 mV (see FIG. \ref{fig4}(right)). Different colors represent different potentials. Then the force due to these patches can be obtained through the runs of simulation. Since the potentials of patches are randomly, we need to repeat the process. Therefore, a mean value and the standard deviation can be obtained on each data point.

\begin{figure*}[htbp]
\centering
\subfigure{%
    \includegraphics[width=3.1in]{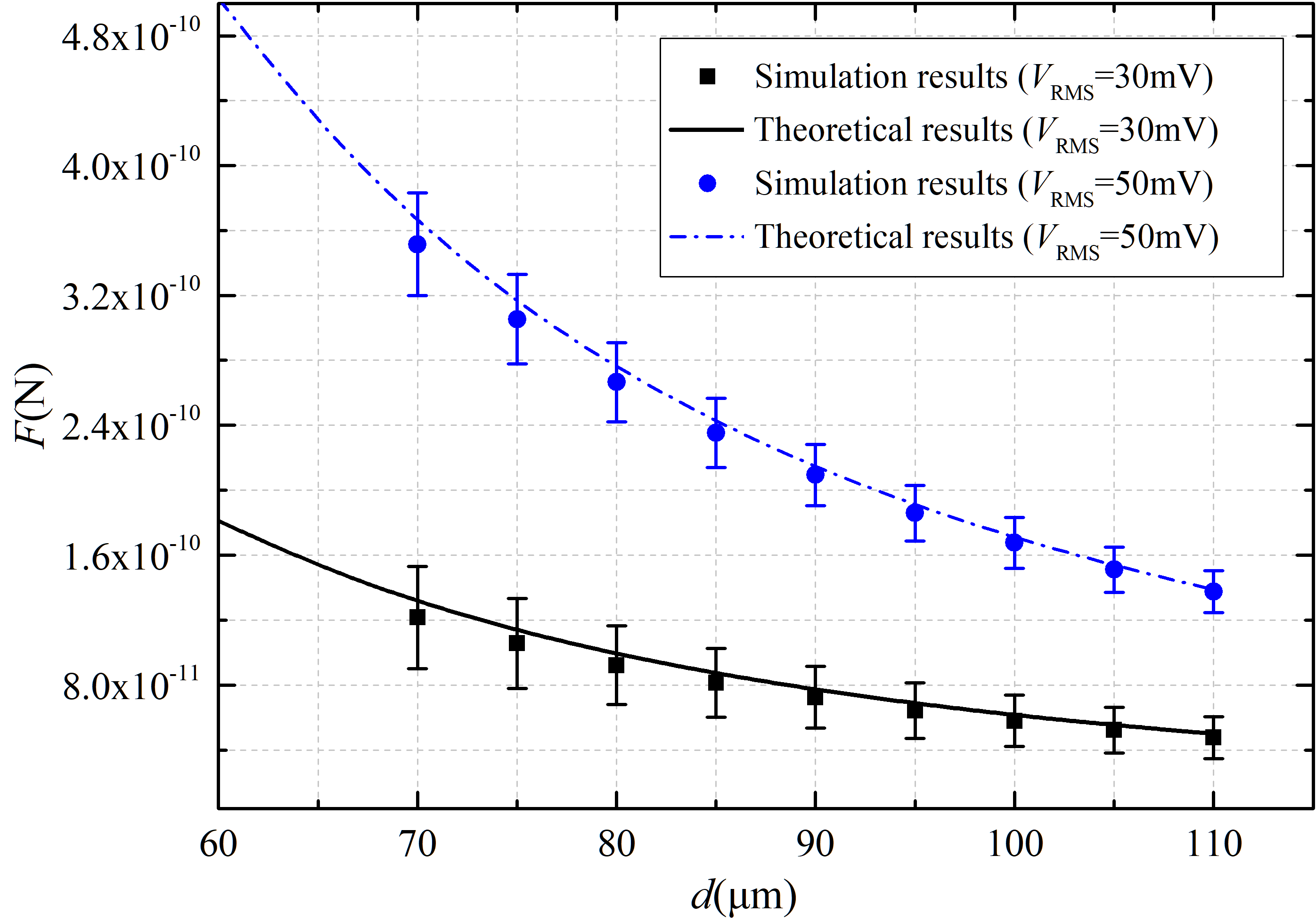}}
\quad
\subfigure{
    \includegraphics[width=3.3in]{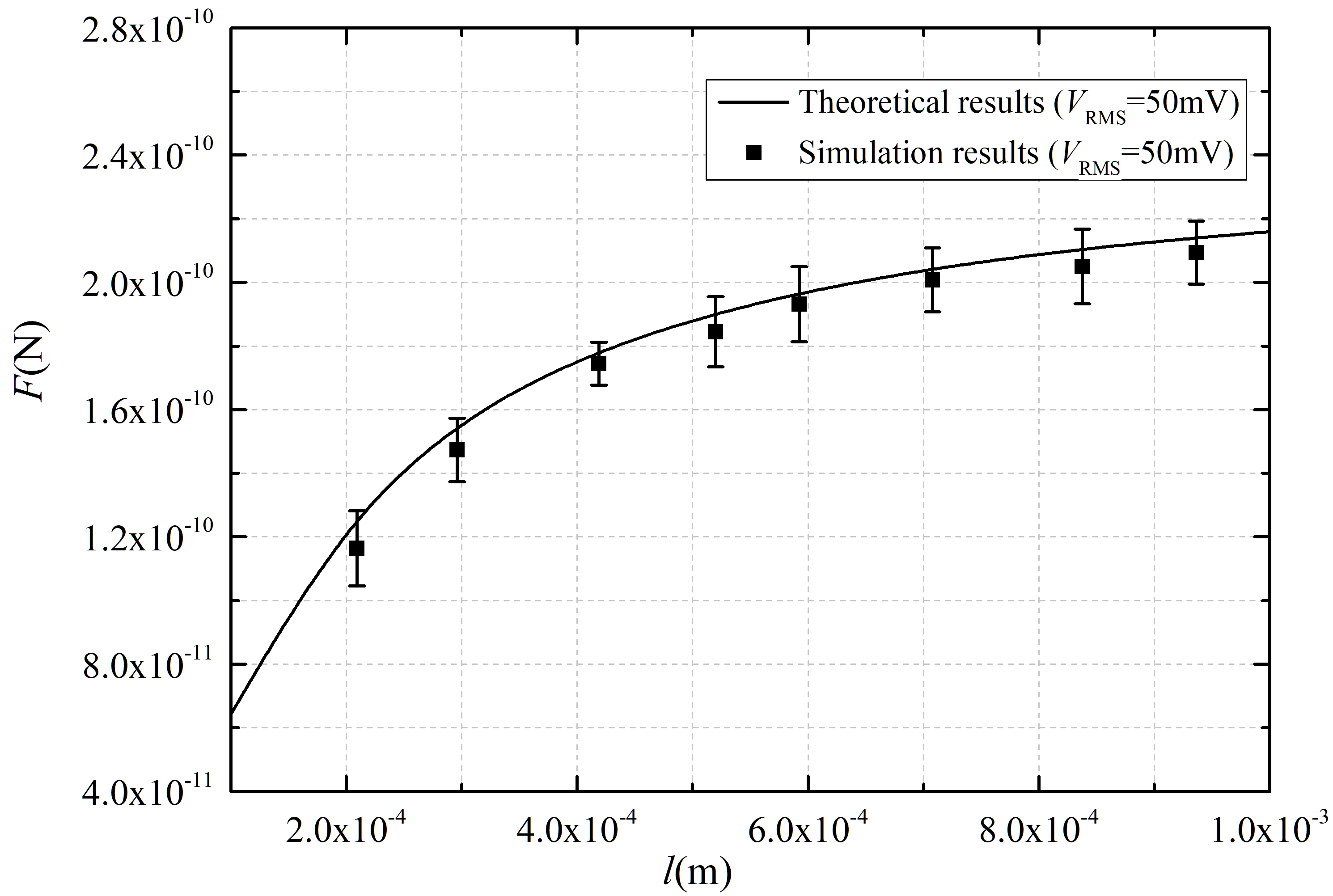}}

\caption{\label{fig5} Comparison between the simulation results and theoretical results for the electrostatic force (left) the electrostatic force between the plates at different separations (with different potential variances 30 mV and 50 mV). (right) the electrostatic force between the plates with different effective patch length, the distance is 90 ${\rm \mu m}$.}
\end{figure*}

We first explore the finite size effects at different separations with 200 patches. As shown in FIG. \ref{fig5} (left), the simulated electrostatic forces (black square and blue circle) are plotted as a function of separations with different potential variances. The simulation results are in agreement with theoretical results (black solid line and blue dotted line) at different separations. The standard deviations are about one-tenth of the mean values. It is worth emphasizing that we only check the autocorrelation function by setting one patch surface. Then we study the finite size effects of different lengths, as shown in FIG. \ref{fig5} (right). The simulation results (black square) also meet well with theoretical results (black solid line). Therefore, we can conclude that the finite size effects of plates are at an acceptable level and the approximation of Eq. (\ref{N16}) is effective. The number of patches is limited by the storage space of computer.

\section{Temporal variation of the surface potential}
In addition to the spatial variation of the surface potential, there also exist the temporal variation of the surface potential \cite{4,9,21,22,23,24,43}. Generally, investigators always use the temporal fluctuation of mean potential to calculate its influence on experiments \cite{4,14,15}. Therefore, we assume that the potentials of surface patches share the same temporal fluctuation and use the temporal fluctuation of mean potential to represent this property. The mechanism of potential fluctuation is not clear until now. Several models have been suggested as underlying mechanism for this fluctuation, such as fluctuating adatomic dipoles and adatom diffusion. Ref. \cite{4} also suspect that the outgassing of particles related electrical effects may be the possible explanation. In this paper, we adopt the adatom diffusion on surface as the cause of this fluctuation. Adatoms that diffuse in and out of the surface change the average work function and lead to fluctuations of the mean potential \cite{44}. The relationship can be written as
\begin{equation}\label{N38}
\delta \bar{V}=\frac{\partial \bar{V}}{\partial \bar{n}}\delta \bar{n},
\end{equation}
where ${\delta \bar{V}}$ and ${\delta \bar{n}}$ are the fluctuations in the density of adatoms and in the work function, respectively. Based on the model in Ref. \cite{24}, we can relate the mean density of adatoms ${\bar{n}(t)}$ to the mean dipole moment density ${\bar{P}(t)}$ by ${\bar{P}(t)=\bar{\mu }\bar{n}(t)}$, where ${\mu}$ is the mean dipole moment. Thus
\begin{equation}\label{N39}
\delta \bar{V}=\frac{{\bar{\mu }}}{{{\varepsilon }_{0}}S}\delta N
\end{equation}
where ${N}$ is the value of the total number of atoms on surface. It is worth emphasizing that the spatial distribution on the potentials have been neglected here. Since we suspect that this distribution may be small in contrast to the variation cause by the model in section 3. Consequently, a spectral analysis for Eq. (\ref{N39}) can be conducted
\begin{equation}\label{N40}
{{S}_{\delta \bar{V}}}(\omega )=\frac{{\bar{\mu }}}{{{\varepsilon }_{0}}A}{{S}_{\delta N}}(\omega )
\end{equation}
Based on the analysis in Ref. \cite{45,46}, it is easy to obtain the correlation function of the number fluctuation of adatoms in a fixed area ${S}$ caused by random diffusion
\begin{equation}\label{N41}
\left\langle \delta N(0)\delta N(\tau ) \right\rangle =\frac{\left\langle {{(\delta N)}^{2}} \right\rangle }{S}\int_{S}{{{d}^{2}}\vec{r}}\int_{S}{{{d}^{2}}{\vec{r}}'}\frac{1}{4\pi D\tau }{{e}^{-\frac{{{\left| \vec{r}-{\vec{r}}' \right|}^{2}}}{4D\tau }}}
\end{equation}
where ${\left\langle {{(\delta N)}^{2}} \right\rangle }$ is the mean square fluctuation of adatoms in ${S}$ and ${D}$ is the diffusion constant. The power spectral density of ${\delta N}$ can be written as
\begin{equation}\label{N42}
S_{\delta N}^{2}(\omega )=2\operatorname{Re}\left[ \int_{0}^{\infty }{\left\langle \delta N(0)\delta N(\tau ) \right\rangle {{e}^{-j\omega t}}dt} \right]
\end{equation}
Combining Eq. (\ref{N41}) and (\ref{N42}), we have
\begin{equation}\label{N43}
S_{\delta N}^{2}(\omega )=\frac{\left\langle {{(\delta N)}^{2}} \right\rangle }{\pi DS}\int_{S}{{{d}^{2}}\vec{r}}\int_{S}{{{d}^{2}}{\vec{r}}'}\text{Ke}{{\text{r}}_{0}}(\left| \vec{r}-{\vec{r}}' \right|\sqrt{\omega /D})
\end{equation}
where ${\text{Ke}{{\text{r}}_{0}}(x)}$ is the zeroth-order Kelvin function. It is hard to obtain a theoretical result for a rectangular surface. But we can approximate the rectangular surface as a circle by using ${S=\pi {{R}^{2}}}$. In this case, Eq. (\ref{N43}) is given by
\begin{equation}\label{N44}
S_{\delta N}^{2}(\omega )=-\left\langle {{(\delta N)}^{2}} \right\rangle \frac{8{{R}^{2}}}{D}\frac{\text{Be}{{\text{r}}_{1}}(u_{p}^{1/2})\text{Ke}{{\text{i}}_{1}}(u_{p}^{1/2})+\text{Be}{{\text{i}}_{1}}(u_{p}^{1/2})\text{Ke}{{\text{r}}_{1}}(u_{p}^{1/2})}{{{u}_{p}}}
\end{equation}
where ${{{u}_{p}}=\omega {{R}^{2}}/D}$ and ${\text{Be}{{\text{r}}_{1}}(x),\text{Ke}{{\text{i}}_{1}}(x),\text{Be}{{\text{i}}_{1}}(x),\text{Ke}{{\text{r}}_{1}}(x)}$ are the Kelvin functions. Therefore, the final form of the fluctuation in mean potential is
\begin{equation}\label{N45}
{{S}_{\delta \bar{V}}}(\omega )=\frac{{\bar{\mu }}}{{{\varepsilon }_{0}}\pi R}{{\left( \frac{8\left\langle {{(\delta N)}^{2}} \right\rangle }{D}\Theta ({{u}_{p}}) \right)}^{1/2}}
\end{equation}
where
\begin{equation}\label{N46}
\Theta ({{u}_{p}})=\frac{\text{Be}{{\text{r}}_{1}}(u_{p}^{1/2})\text{Ke}{{\text{i}}_{1}}(u_{p}^{1/2})+\text{Be}{{\text{i}}_{1}}(u_{p}^{1/2})\text{Ke}{{\text{r}}_{1}}(u_{p}^{1/2})}{{{u}_{p}}}
\end{equation}
Thus, we can plot ${\Theta ({{u}_{p}})}$ as a function of ${u_p}$, as shown in FIG. \ref{fig4}. Note that two limiting cases can be derived: for the low-frequency part, the spectrum is
\begin{equation}\label{N47}
{{S}_{\delta \bar{V}}}(\omega \to 0)=\frac{{\bar{\mu }}}{{{\varepsilon }_{0}}\pi R}{{\left( \frac{\left\langle {{(\delta N)}^{2}} \right\rangle }{D}\ln (\frac{D}{\omega {{R}^{2}}}) \right)}^{1/2}}.
\end{equation}
for the high-frequency part, the spectrum is
\begin{equation}\label{N48}
{{S}_{\delta \bar{V}}}(\omega \to \infty )=\frac{{\bar{\mu }}}{{{\varepsilon }_{0}}\pi {{R}^{2}}}{{\left( \frac{{{8}^{1/2}}{{D}^{1/2}}\left\langle {{(\delta N)}^{2}} \right\rangle }{R}{{\omega }^{-3/2}} \right)}^{1/2}}.
\end{equation}

\begin{figure*}[htbp]
\includegraphics[width=0.45\textwidth]{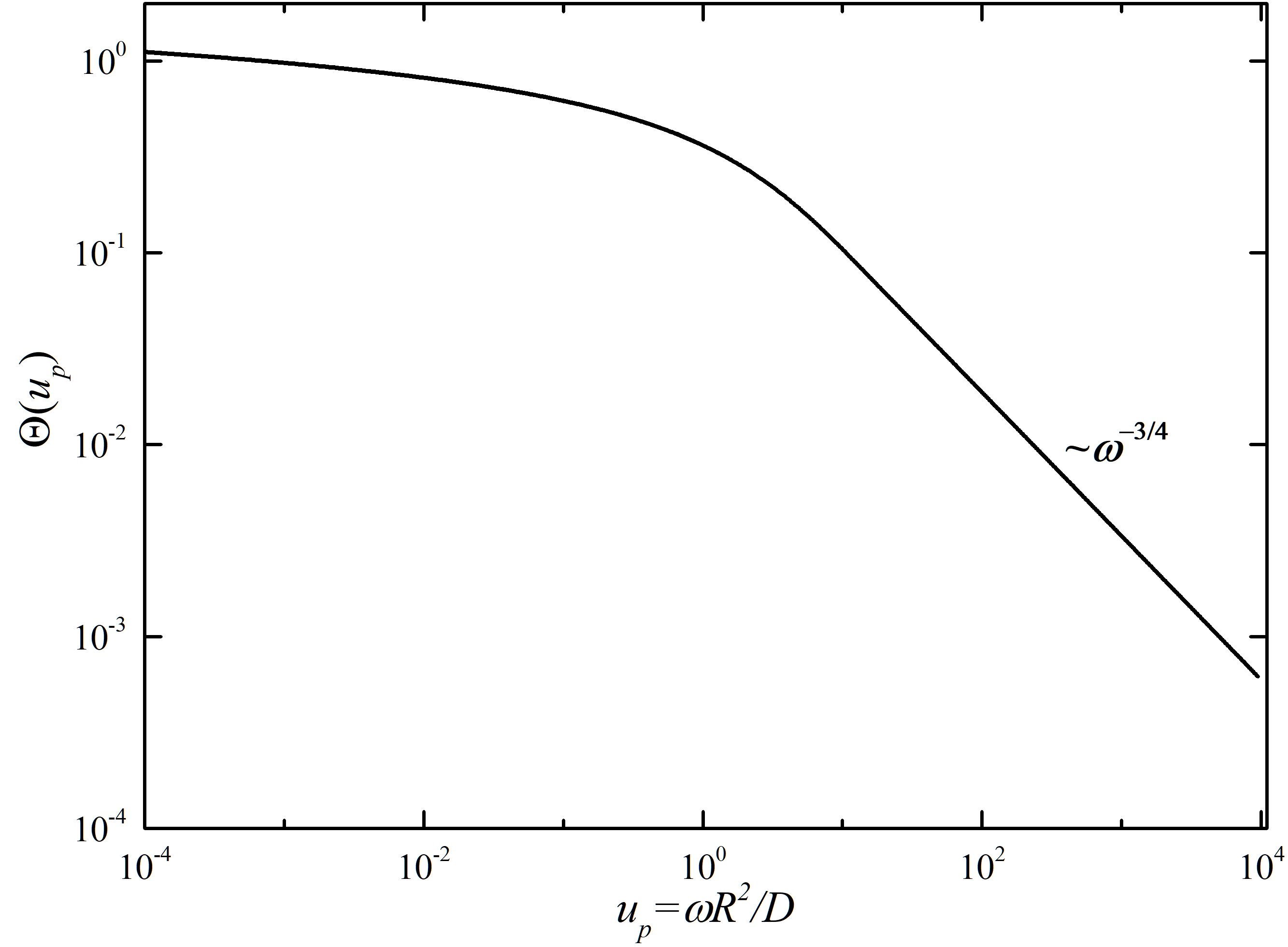}
\caption{Dependence of the spectral function ${\Theta ({{u}_{p}})}$ on the scaled frequency ${u_p}$.}
\label{fig6}
\end{figure*}

\section{Experimental verification of the residual electrostatic effect}
In the previous section, we have discussed the model for the random patch force. Now we conduct an experiment to verify these models. This torsion balance experiment is specifically designed to investigate residual electrostatic effects. In this experiment, an I-shaped pendulum with a mass of 13 g, was suspended facing to the membrane by an 80-mm-long, 25-${\rm \mu m}$-diameter tungsten fiber. A ${20 \times 20 \times 0.005 \; \rm mm^3}$ conducting membrane is placed at one side of the pendulum. The pendulum and the membrane were all gold coated. A schematic of our apparatus is shown in FIG. \ref{fig5}. The separation between the pendulum and membrane can be adjusted between 0 and 200 ${\rm \mu m}$ within 1 ${\rm \mu m}$ in accuracy. The sensitivity of the closed-loop pendulum was calibrated synchronously
by a rotating copper cylinder. The apparatus was housed inside a vacuum chamber with a pressure of approximately ${10^{-5}}$ Pa. More details of this experimental design can be found in Ref. \cite{47}.

\begin{figure*}[htbp]
\includegraphics[width=0.6\textwidth]{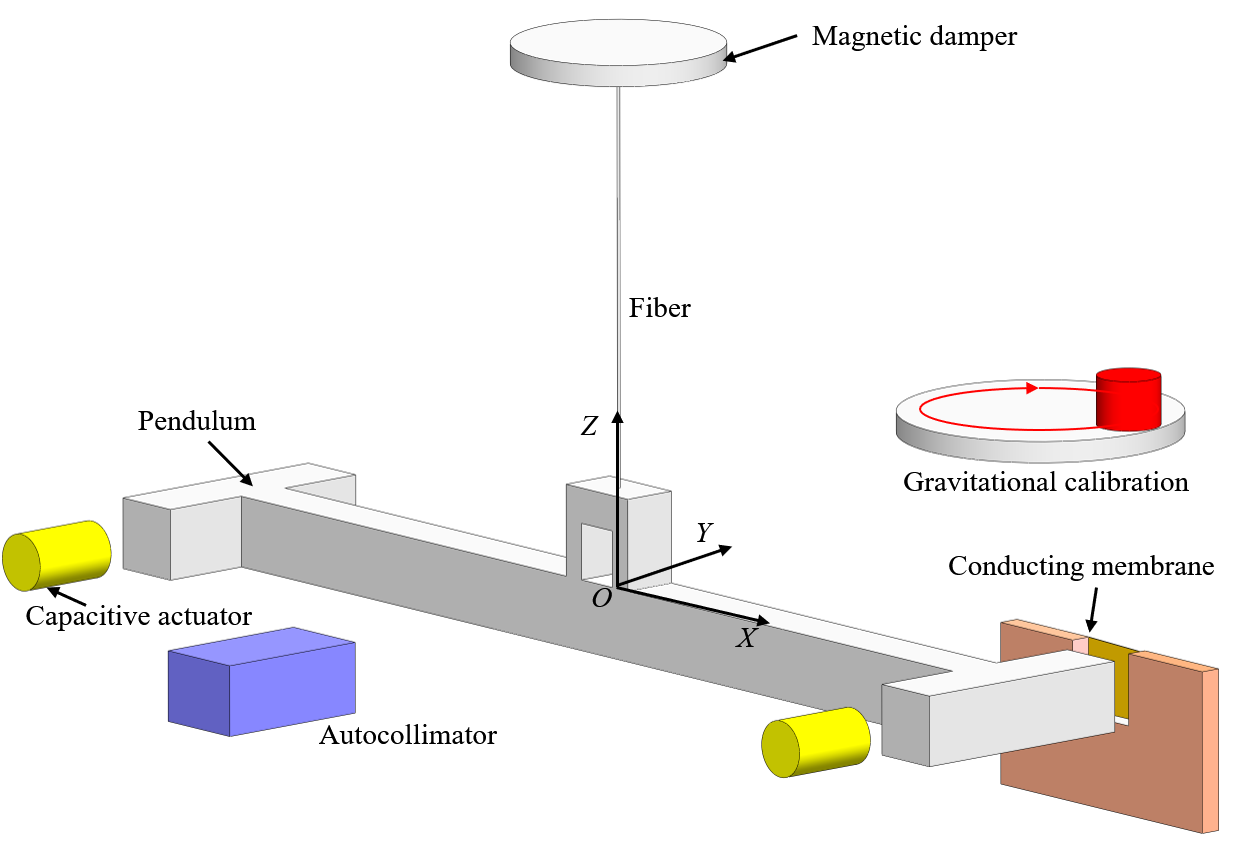}
\caption{Schematic drawing of the experimental apparatus.}
\label{fig7}
\end{figure*}

In the ${\rm \mu m}$ range, the electrostatic disturbance is the dominant noise source. In order to maintain the stability of the separation between the pendulum and membrane, a proportional-integral-differential (PID) electrostatic feedback control system was used. Therefore, we can obtain the torque exerted on the pendulum by using the feedback voltage. In addition, the contact potential differences between the pendulum and membrane were compensated by applying a voltage on the membrane. Usually, we think that the mean patch force can be eliminated by applying compensation voltage and the random patch force still exist after compensating. Therefore, it is reasonable to conclude that the residual electrostatic torque in this experiment is mainly caused by the random patch potentials. As already stated, the patch potentials vary with time. In order to avoid the influence of potential fluctuation, the surface potential of each membrane needs be compensated for each data run. The measurement period can not be so long.

By processing the experimental data, we have obtained the residual electrostatic gradient at different separations (points with error bars), as shown in FIG. \ref{fig6} (left). These points show a large residual electrostatic gradient exist in the ${\rm \mu m}$ range. We convert Eq. (\ref{N35}) into the torque form by a moment arm 47.5 mm and then fit these results. A best fit is achieved and gives ${V_{rms}^{2}={{(20\;\text{mV})}^{2}}}$, ${w = 0.4}$ mm and ${\chi \approx -3}$ (black solid line). We note that the predicted mean patch size is larger than ${a}$ for the whole range of distances conducted in the experiment (6.0-40.0) ${\rm \mu m}$. Usually, typical values of patch size reported by experiments are in the range 10 nm to 1 ${\rm \mu m}$ and is much smaller than our result. This big patch size can be interpreted as two possible reasons: one is that the absorption of contaminants in surface alter the patch size. Another is that the interaction of the patches on two surfaces. Since we can not model the cross correlations between the patches on different plates, we always ignore this effect. This approximation may be valid when the area of one surface is relative smaller than another, while invalid when the areas are almost identical.

\begin{figure*}[htbp]
\centering
\subfigure{%
    \includegraphics[width=3.05in]{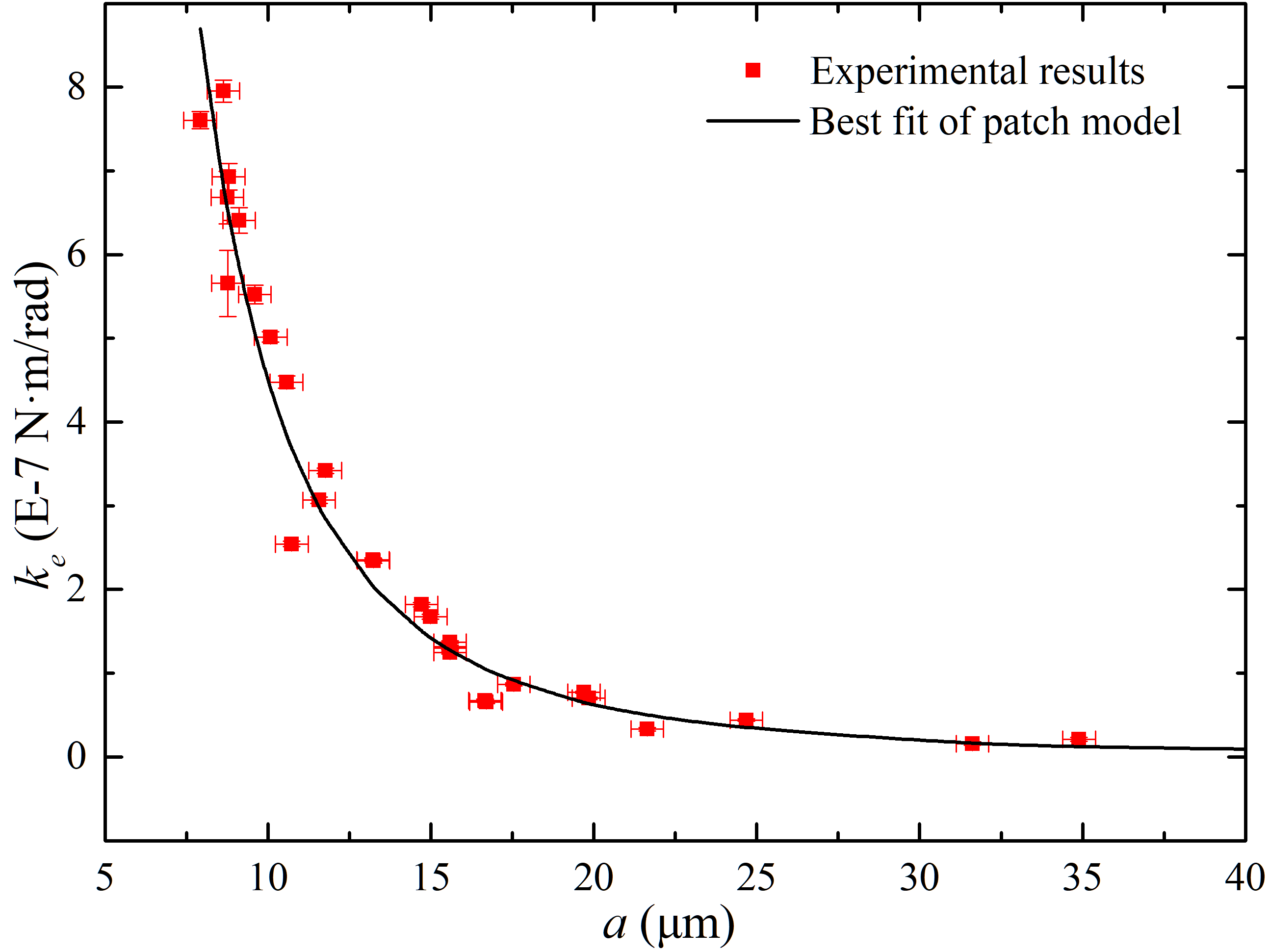}}
\quad
\subfigure{
    \includegraphics[width=3.15in]{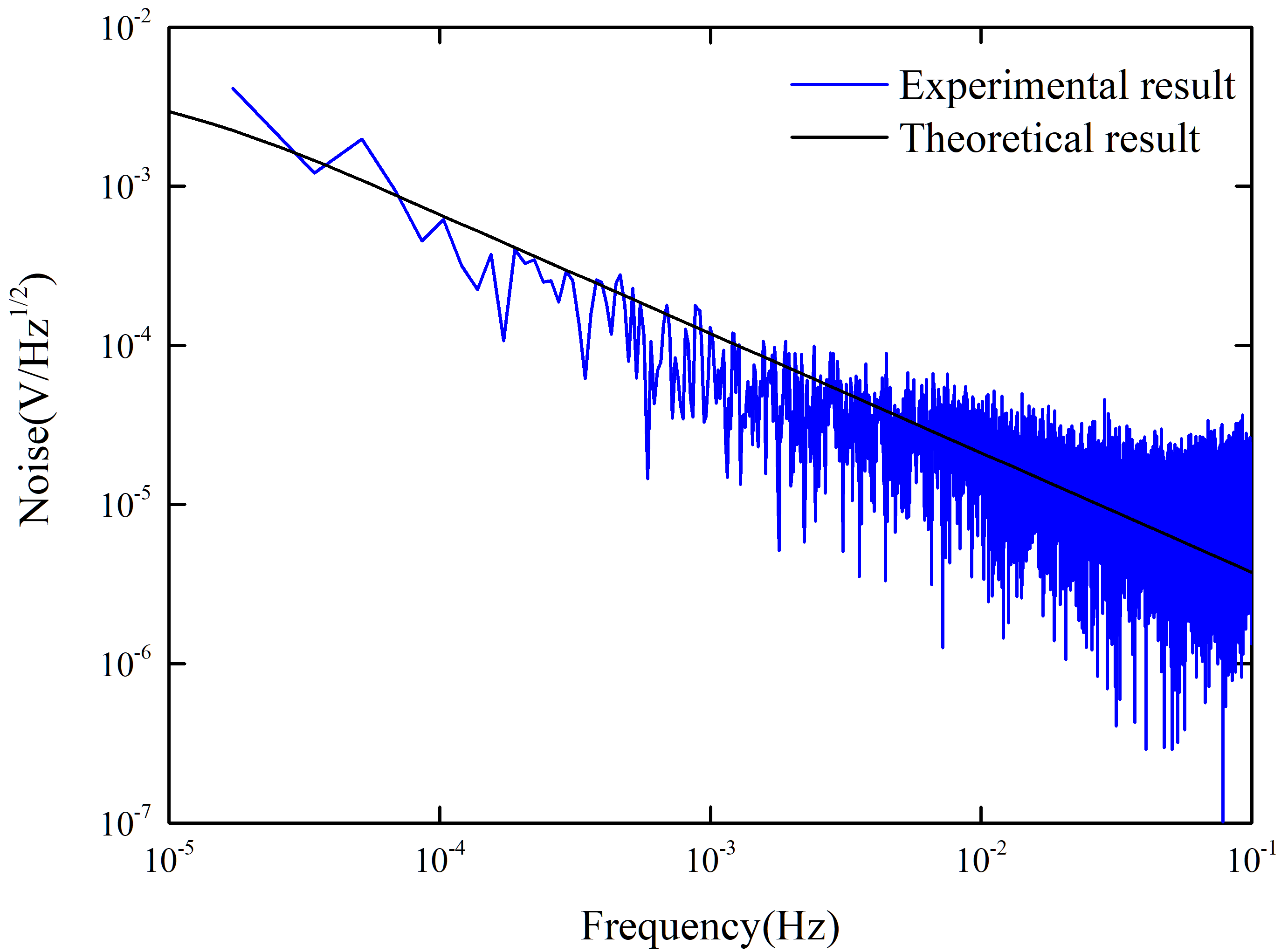}}

\caption{(left) Comparison of the residual random patch torque gradient between the experimental data (points with error bars) and the best fit of the patch torque gradient within the two-dimensional quasi-local model of section 3 (black solid line). (right) Comparison of the voltage fluctuation (blue solid line) and the theoretical prediction by using Eq. (39) (black dashed line).}
\label{fig8}
\end{figure*}

In order to verify the model in Sec. IV, we compared the theoretical frequency dependence of mean potential with the experimental results provided by HUST group \cite{6,7}. This experiment measured the temporal and spatial variation of surface potential by using a torsion pendulum and a scanning conducting probe. The areas of the pendulum and the scanning probe are
${100 \times 40 \; \rm mm^2}$ and ${5 \times 5 \; \rm mm^2}$, respectively. The distance between the test mass and the probe is 100 ${\rm \mu m}$. More details can be found in Refs. \cite{6} and \cite{9}. A typical potential variation of the test mass is shown in Fig. \ref{fig8} (right). The spectrum rises as about ${1/f}$ below 0.01 Hz and is about ${\rm 50 \mu V/Hz^{1/2}}$ at 1 mHz. We then plot a theoretical line by using Eq. (\ref{N45}), and some typical parameters are used here, such as ${\mu =5 \; \text{D,}\left\langle {{(\delta N)}^{2}} \right\rangle ={{({2.5 \times {10}^{13}})}^{2}}, \; D=5 \times {{2.5}^{-10}} \; {{\text{m}}^{2}}{{\text{s}}^{-1}}.}$ This model predicts a ${1/f^{3/4}}$ scaling in the frequency range from ${10^{-5}}$ Hz to ${1}$ Hz and predicts a flatten spectrum below ${10 \rm \mu m}$ Hz. From FIG. \ref{fig6} (right), we see that the experimental result meets well with theoretical prediction from ${10^{-5}}$ Hz to ${10^{-2}}$ Hz and the experimental noise is larger than expected in the high-frequency part. Therefore, we suspect that the adatom diffusion may be the possible mechanism for the voltage fluctuation.
\section{Applications}
Generally, we are not interested in the DC electrostatic force (or torque), but in the force fluctuating at the target frequency. The residual electrostatic effect produces force noise in two ways. First, fluctuations in distance will multiply the spring constant of the electrostatic interaction to produce force noise. Based on Eq. (\ref{N35})
\begin{equation}\label{N49}
{{S}_{{{F}_{\text{res,random}}}(\delta a)}}(\omega )={{K}_{e}}{{S}_{\delta a}}(\omega )=\frac{4{{\varepsilon }_{0}}SV_{rms}^{2}}{\pi {{a}^{3}}}{{(\frac{a}{w})}^{\chi (a/w)+3}}{{S}_{\delta a}}(\omega ),
\end{equation}
where ${\delta a\ll a}$ has been used and ${S_{\delta a}^{2}(\omega )=\int{d\tau \left\langle \delta a(t)\delta a(t+\tau ) \right\rangle {{e}^{-j\omega \tau }}}}$ is the displacement noise.

Second, any temporal variation of the patch potential will also multiply the force gradient of voltage to produce force noise. By assuming that the potentials of surface patches share the same temporal fluctuation and using the temporal fluctuation of mean potential to represent this property. We obtain
\begin{equation}\label{N50}
{{S}_{{{F}_{\text{res,mean}}}(\delta \bar{V})}}(\omega )=\left| \frac{\partial F}{\partial ({{V}_{\text{rms}}})} \right|{{S}_{\delta \bar{V}}}(\omega )=\frac{4{{\varepsilon }_{0}}S{{V}_{\text{rms}}}}{\pi {{a}^{2}}}{{(\frac{a}{w})}^{\beta (a/w)+2}}{{S}_{\delta \bar{V}}}(\omega ).
\end{equation}
where ${{{S}_{\delta \bar{V}}}(\omega )}$ is determined by Eq. (\ref{N45}).
\subsection{Applications to the experiment of testing ISL}
We now apply the models to analyze the residual electrostatic effects existed in the ISL experiment by the HUST (Huazhong University of Science and Technology) group. In fact, the experimental apparatus and environmental condition in Ref. \cite{15} are similar to our experiment in Sec. V. The mean patch potential of each membrane was compensated to equipotential with the pendulum for each data run. Based on which, it is reasonable to assume that the distribution of patch size is also similar. In this condition, the torque gradient of distance can be written as
\begin{equation}\label{N51}
{{K}_{e,T}}\approx \frac{8{{\varepsilon }_{0}}SV_{rms}^{2}{{L}^{2}}}{\pi {{a}^{3}}},
\end{equation}
where the gradient is summed over two sides of the pendulum, ${a}$ is the separation between the test mass and membrane (about 90 ${\rm \mu m}$), and ${L}$ is the distance between the fiber and the center of the test mass (about 38 mm). According to the result of Ref. \cite{15}, the torque gradient of distance is about ${\rm 2.6 \pm 0.4 \times 10^{-8}}$ Nm/rad, which can be obtained by using ${{{V}_{rms}}=50}$ mV and ${w= }$ 0.4 mm, respectively. The tiny vibration of the shielding membrane is about 0.1 nrad and the disturbance of the torque from this vibration is estimated as <${\rm 0.3 \times 10^{-17}}$ Nm.

For the torque gradient of voltage, we have
\begin{equation}\label{N52}
{{K}_{v,T}}\approx \frac{8{{\varepsilon }_{0}}S{{V}_{\text{rms}}}L}{\pi {{a}^{2}}}.
\end{equation}
Inserting the parameters into this equation, we can estimate the value as ${\rm 9.3 \times 10^{-10}}$ Nm/V. If we assume that the noise floor of voltage is less than 50 ${\rm \mu V/Hz^{1/2}}$ around several millihertz. The voltage noise introduces a torque noise of ${\rm 4.65 \times 10^{-14} \; Nm/Hz^{1/2}}$ , which is about 25 times larger than the thermal noise ${\rm 2 \times 10^{-15} \; Nm/Hz^{1/2}}$ of the torsion balance with ${\rm Q\approx 2500}$. After an integration of 10 days, the disturbance of the torque from voltage noise is estimated as ${\rm 5 \times 10^{-17}}$ Nm and is almost identical to the experimental sensitivity.
\subsection{Applications to LISA}
The patch-field related effects have been recognized as important error sources for LISA. As already stated, the surface potential can be divided into the contact potential differences and random patch potential. We consider the situation that the sensing voltage are not applied on the electrodes. The first noise term is the stiffness due to random patch potentials. In their former disturbance requirement, they used a function based on the model discussed in Ref. \cite{11}, the stiffness formula is \cite{4,35,36,38}
\begin{equation}\label{N53}
{{K}_{e}}=\gamma \frac{{{\varepsilon }_{0}}SV_{rms}^{2}}{m{{a}^{3}}}.
\end{equation}
where ${\gamma}$ is a dimensionless constant of order 1. This formula was derived under the assumption of a sharp-off model and the worst case was selected. In Ref. \cite{36,38}, ${\gamma}$ and ${{{V}_{rms}}}$ are assumed to be 1.8 and 100 mV, respectively. After substituting the geometric parameters into Eq. (\ref{N53}), we obtain ${{{K}_{e}}=2.65\times {{10}^{-9}} \; {{\text{s}}^{-2}}}$ in ${x}$ axis (4 electrodes), where ${S=14.5\times 36 \; \text{m}{{\text{m}}^{2}}}$, ${m=1.96 \; \text{kg}}$ and ${a=4}$ mm are used. As mentioned before, this model does not have a suitable physical explanation. We then apply the model in this paper to give an estimation. Since we have no information about the patch size in this experiment, we can obtain the stiffness result as a function of ${w}$ by using the parameters mentioned before, as shown in FIG. \ref{fig7} (${{{V}_{rms}}}$ = 100 mV). The stiffness increases rapidly with the increasing of ${w}$ and reaches the maximum value ${3\times {{10}^{-9}} \; {{\text{s}}^{-2}}}$ with ${w = 0.1}$ m. However, this big size can not be realized. We choose a 3 mm-scale patches and obtain a stiffness value about ${8.47\times {{10}^{-10}} \; {{\text{s}}^{-2}}}$. Therefore, this kind of stiffness meets the requirement of the maximum total parasitic coupling. The second noise term is the acceleration noise induced by the voltage noise and random patch potentials. The acceleration gradient of voltage is about ${2.03\times {{10}^{-11}} \; \text{m}{{\text{s}}^{-2}}/\text{V}}$ according to Eq. (\ref{N50}). Similarity, the acceleration disturbance from this noise is estimated as ${1.02\times {{10}^{-15}} \; \text{m}{{\text{s}}^{-2}}/\sqrt{\text{Hz}}}$ with a noise floor of voltage ${5 \times {{10}^{-5}} \; \text{V}/\sqrt{\text{Hz}}}$.

\begin{figure*}[htbp]
\includegraphics[width=0.50\textwidth]{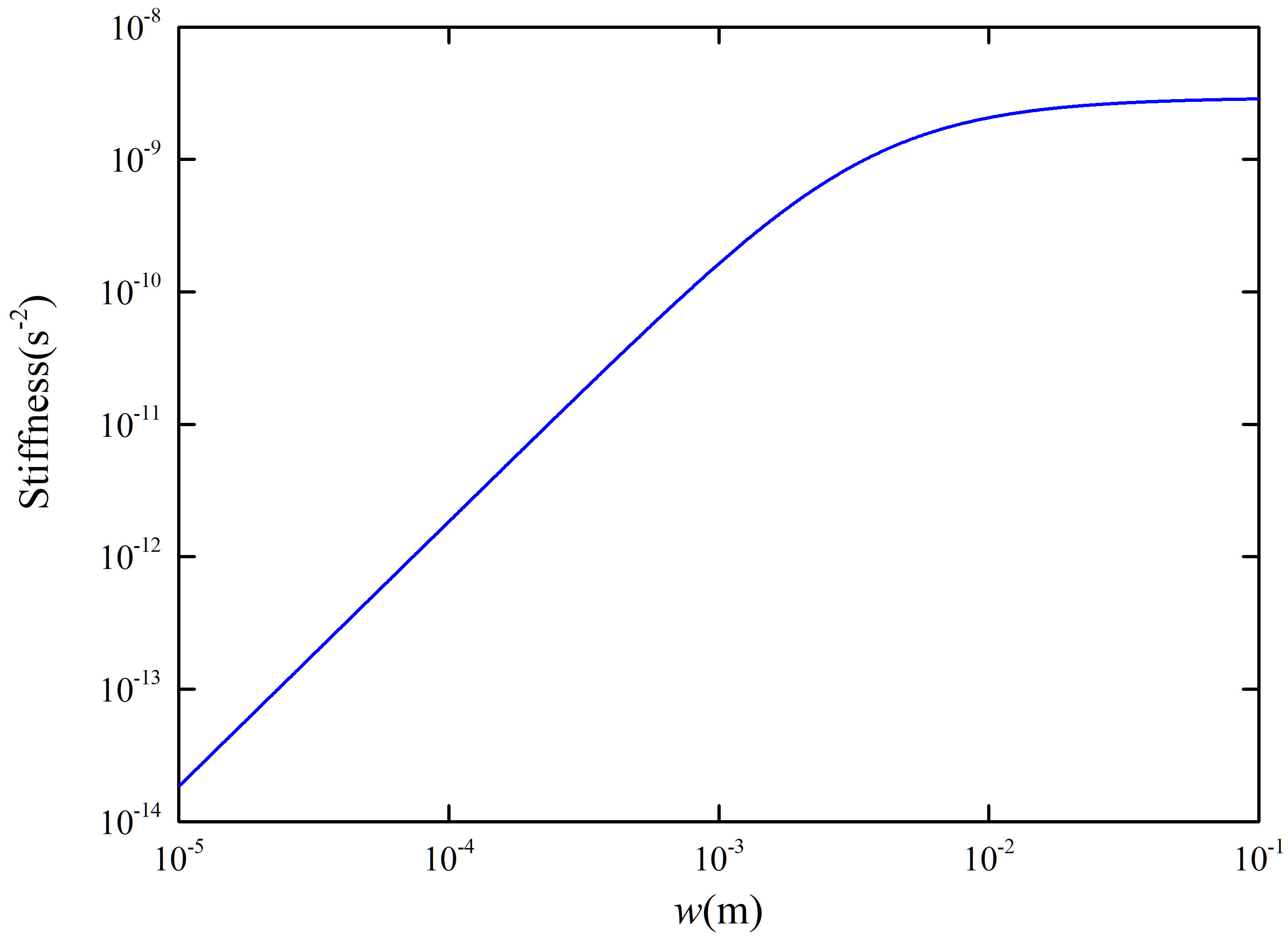}
\caption{Dependence of the stiffness on the patch length ${w}$.}
\label{fig9}
\end{figure*}

The third and fourth noise terms arise from the interactions between the free charge on the test mass (TM) and the contact potential differences between opposing electrodes \cite{48}. In fact, the influence of these terms have been verified by using an electrostatic measurement made on board the LISA Pathfinder \cite{49}. Meanwhile, we focus on the random patch potential rather than the contact potential differences. We hence adopt their results directly, which show that the level of charge-induced acceleration noise on single TM (including the couping with mean patch potential difference) is about 1.0 fm ${\rm s^{-2} \; Hz^{-1/2}}$ across the 0.1-100 mHz frequency band \cite{49}.
\section{Discussion}
To summarize, we give a full analysis for the patch effect between closely spaced surfaces, including theoretical modeling, numerical analysis and experimental verification. For the spatial potential variation, we have further improved the quasi-local correlation model, and have obtained a rigorous formula based on the Poisson statistic of patch sizes. A cleaner relationship between the force and the patch size are given. The experimental results show a 0.4 mm effective patch size, which is larger than the empirical values. This comparison indicates a big difference between the patch force of two closely spaced surfaces and the electric filed of one surface. For the temporal potential variation, we have used an adatom diffusion model to describe the voltage fluctuating. A good agreement between the theoretical prediction and experimental data from ${10^{-5}}$ Hz to ${10^{-3}}$ Hz shows that this mechanism may be the possible explanation for the fluctuation. Finally, we have analyzed the noises induced by random patch potentials for HUST-2020 ISL experiment and LISA with a revised quasi-local correlation model. These results show that these effects are potentially among the largest in these two experimental budgets.

The challenges of forthcoming studies may be stated as follows. First, it is important to investigate the interaction between the spatial potential variation and temporal potential variation. This problem is difficult to explore because of the unknown origin of patch potential. Second, it would also be important to study the cross correlations between the patches on different plates. We are currently performing finite element analysis for this correlation. Third, it is necessary to perform Kelvin probe force microscopy measurement with different probe size to measure the spatial variations of the potential to confirm the hypothesis of the revised quasi-local correlation model. In fact, Garrett ${et \, al}$ have performed a measurement to the correlation between the patch potentials over a surface \cite{8}. Their result shows that, there still exist a strong correlation between two long spaced patches. Therefore, a further investigation to explore this correlation should be carried out. Finally, a comparison between different mechanisms for localized field fluctuations is needed. Anyway, to obtain a higher experimental sensitivity, we should prepare a cleaner surface with smaller patch size, which can possible be realized by using some preparation techniques, like the technologies of template-stripping and annealing \cite{8}.
\\

\section*{ACKNOWLEDGMENTS}
We are grateful to Hang Yin, and Chi Song for having kindly provided experimental data and information needed to analyze them. This work is supported by National Key R\rm \&D Program of China (Grant No. 2020YFC2200500), the National Natural Science Foundation of China (Grant Nos. 12150012, 11805074, and 11925503), Guangdong Major Project of Basic and Applied Basic Research (Grant No. 2019B030302001), and the Fundamental Research Funds for the Central Universities, HUST: 2172019kfyRCPY029.

\end{document}